\documentclass[acmlarge]{acmart}
\AtBeginDocument{%
  }

\setcopyright{acmlicensed}
\copyrightyear{2018}
\acmYear{2018}
\acmDOI{XXXXXXX.XXXXXXX}

\acmISBN{978-1-4503-XXXX-X/18/06}

\usepackage{algorithm}
\usepackage{algpseudocode}
\usepackage{adjustbox}
\usepackage[utf8]{inputenc}
\usepackage{url} % For formatting URLs
\usepackage{hyperref} % For creating clickable links
\usepackage{booktabs}
\usepackage{graphicx}
\usepackage{lscape}
\usepackage{wrapfig}
\usepackage{enumerate}
\usepackage{graphicx}
\usepackage{caption}
\usepackage{subcaption}
\usepackage{array}
\usepackage{tabularx}
\usepackage{wrapfig}
\usepackage{subcaption}
\usepackage{soul}
\usepackage{multirow}
\begin{document}

\title{Are Anxiety Detection Models Generalizable? A Cross-Activity and Cross-Population Study Using Wearables}
%Investigating Generalizability of Anxiety Detection Models using Wearable Sensors for anxious activities and Participants}

\author{Nilesh Kumar Sahu}
\email{nilesh21@iiserb.ac.in}
\orcid{0000-0003-1675-7270}
\affiliation{%
  \institution{Indian Institute of Science Education and Research Bhopal (IISERB)}
  \city{Bhopal}
  \state{Madhya Pradesh}
  \country{India}
}

\author{Snehil Gupta}
\email{snehil.psy@aiimsbhopal.edu.in}
\orcid{0000-0001-5498-2917}
\affiliation{%
  \institution{All India Institute of Medical Sciences Bhopal}
  \city{Bhopal}
  \state{Madhya Pradesh}
  \country{India}
}

\author{Haroon R Lone}
\email{haroon@iiserb.ac.in}
\orcid{0000-0002-1245-2974}
\affiliation{%
  \institution{Indian Institute of Science Education and Research Bhopal (IISERB)}
  \city{Bhopal}
  \state{Madhya Pradesh}
  \country{India}
}

\renewcommand{\shortauthors}{Sahu et al.}
\renewcommand{\shorttitle}{Investigating Generalizability of Anxiety Detection Models}



\begin{abstract}

Anxiety-provoking activities, such as public speaking, can trigger heightened anxiety responses in individuals with anxiety disorders. Recent research suggests that physiological signals, including electrocardiogram (ECG) and electrodermal activity (EDA), collected via wearable devices, can be used to detect anxiety in such contexts through machine learning models. However, the generalizability of these anxiety prediction models across different activities and diverse populations remains underexplored—an essential step for assessing model bias and fostering user trust in broader applications. To address this gap, we conducted a study with 111 participants who engaged in three anxiety-provoking activities.

Utilizing both our collected dataset and two well-known publicly available datasets, we evaluated the generalizability of anxiety detection models within participants (for both same-activity and cross-activity scenarios) and across participants (within-activity and cross-activity). In total, we trained and tested more than 3348 anxiety detection models (using six classifiers, 31 feature sets, and 18 train-test configurations). Our results indicate that three key metrics—AUROC, recall for anxious states, and recall for non-anxious states—were slightly above the baseline score of 0.5. The best AUROC scores ranged from 0.62 to 0.73, with recall for the anxious class spanning 35.19\% to 74.3\%. Interestingly, model performance (as measured by AUROC) remained relatively stable across different activities and participant groups, though recall for the anxious class did exhibit some variation.

This study sheds light on the generalizability of anxiety detection models across similar and dissimilar activities, both within and between participant groups. Moreover, the datasets used in this study come from two different subcontinents—the Global South and the Global North—offering valuable insights for developing robust anxiety detection systems and enriching the existing literature.
% This study sheds light on the generalizability of anxiety detection models across similar and dissimilar activities, both within and between participant groups. Our findings contribute valuable insights toward developing robust anxiety detection systems, enhancing the understanding of the existing literature.
\end{abstract}


\maketitle
%\clearpage
\section{Introduction}

Anxiety is a common mental health problem affecting 301 million people worldwide \cite{Anxietyd41:online}. Anxiety disorders are characterized by excessive fear and worry, with individuals often experiencing intense apprehension about specific situations, such as social interactions. Symptoms of anxiety disorders can include physical manifestations, like increased heart palpitations, sweating etc., behavioral symptoms, such as irritability, restlessness, etc., and cognitive challenges, such as difficulty concentrating \cite{Anxietyd41:online}. Thus, individuals with anxiety disorders avoid anxiety-inducing situations or endure them with considerable distress \cite{salekin2018weakly}.
%Furthermore, anxiety disorders increase the risk of comorbid conditions, including depression, substance abuse disorders, and suicidal tendencies \cite{Anxietyd41:online}.

Currently, there is no objective measure for the severity of anxiety, so mental health professionals continue to rely on traditional gold-standard methods: clinical interviews and self-reported assessments \cite{rashid2020predicting}. However, clinical interviews are time-consuming and susceptible to human judgment biases, while self-reports are affected by the client’s recall limitations and potential biases \cite{rashid2020predicting, salekin2018weakly}. Both methods also require clients to be motivated to openly share their thoughts, feelings, and experiences with professionals or in self-report questionnaires.

The ubiquitous nature of mobile sensing has proven valuable for capturing insights into mental health. Existing research has utilized passively sensed data, such as accelerometer, pedometer, activity, GPS,  audio, etc., to assess and predict individual mental health states \cite{wang2014studentlife, rashid2020predicting, wang2023detecting,yu2023semi,mishra2020evaluating, xu2023globem, meegahapola2023generalization, kammoun2023understanding, shaukat2021detecting, di2021smartphone}. Additionally, physiological data from passively sensed signals like photoplethysmography (PPG) and electrodermal activity (EDA) are being investigated as markers for mental health disorders \cite{mishra2020evaluating, sano2018identifying}. The passively collected data allows mental health professionals and researchers to better understand the onset of mental health disorders and how individuals react to various situations \cite{pillai2024investigating}.

Existing research on passively sensed data for anxiety detection, conducted by the UbiComp community in both controlled and uncontrolled environments, has shown promising evaluation metrics \cite{rashid2020predicting, salekin2018weakly}. However, the generalizability of these methods to new, unseen data remains largely unexplored, especially across varied participant groups, cultural contexts, anxiety-provoking situations, and sensor devices. Assessing the generalizability of anxiety detection methods is critical to ensure that: (i) models trained on narrow or homogeneous datasets do not inherit biases that could compromise their effectiveness across diverse populations and anxiety contexts; (ii) models are resilient to sensor noise caused by factors like sensor degradation or software updates; and (iii) the technology gains user trust, fostering broader adoption for mental health detection. To our knowledge, only three studies—GLOBEM \cite{xu2023globem}, Mood (involving participants from eight countries) \cite{meegahapola2023generalization}, and a study on suicidality \cite{pillai2024investigating}—have examined aspects of generalizability, though they focused on human modeling, mood inference, and suicidality, respectively. However, no study to date has specifically explored the generalizability of anxiety detection models.

Collecting data via wearables or smartphones over extended periods is challenging due to high costs and participant attrition \cite{yu2023semi}. Additionally, gathering data from diverse participant groups across multiple geographic locations requires separate ethical approvals specific to each site. This process is complicated by (i) varying ethical standards and regulations, (ii) distinct data privacy laws and cross-border data-sharing agreements, (iii) cultural sensitivities, and (iv) logistical and administrative complexities. For example, the study by Meeghahapola et al. \cite{meegahapola2023generalization} collected data from eight sites across different countries, each with its own ethics requirements for data integrity and participant recruitment. Given these challenges, publicly available datasets offer a practical alternative for testing the generalizability of anxiety detection methods. Therefore, in this work, we explore the generalizability of anxiety detection models using our in-house dataset from the Global South alongside two well-known publicly available datasets from the Global North, which were collected from diverse geographic locations. Our approach addresses the following research questions related to generalizability.

\begin{itemize}
\item \textbf{RQ1:} Can anxiety prediction models generalize across different anxiety-inducing activities? Specifically, if a model is trained on data collected from one anxiety-provoking activity, will it accurately predict anxiety when applied to a different anxiety-provoking activity? This question assesses the \textit{cross-activity generalizability} of anxiety prediction models.

\item \textbf{RQ2:} Can anxiety prediction models generalize across different participant groups? If a model is trained on data from one participant group in a specific location, will it accurately predict anxiety in a different participant group from another location when both groups undergo the same anxiety-inducing activity? This question examines the \textit{cross-participant generalizability} of anxiety prediction models.

\end{itemize}

To address \textbf{RQ1}, we designed and conducted a study involving 111 participants who engaged in three distinct anxiety-inducing activities. During each activity, we passively collected physiological data—specifically ECG and EDA signals—using wearable devices. After each activity, participants completed a five-item questionnaire to self-rate their anxiety levels. In our data analysis, we assessed the performance of five machine learning models on predicting anxiety based on biobehavioral data, including both physiological measurements and self-reports. We specifically compared model evaluation metrics to understand how a model's accuracy varies when trained and tested on the same activity (\textit{within-activity}) versus across different activities (\textit{cross-activity}). The within-activity model served as a baseline, while the cross-activity analysis offered insights into each model’s generalizability.
In the within-activity analysis, we used 5-fold cross-validation and achieved the best AUROC of 0.82, with a recall of 72.11\% for the anxious class and 73.68\% for the non-anxious class. 
 In the cross-activity analysis, AUROC values ranged from 0.59 to 0.62, which were lower than the within-activity performance. This raises concerns about cross-activity generalization, even with the same participants involved in cross-activities.

\begin{figure*}
    \centering
    \includegraphics[width=1\linewidth]{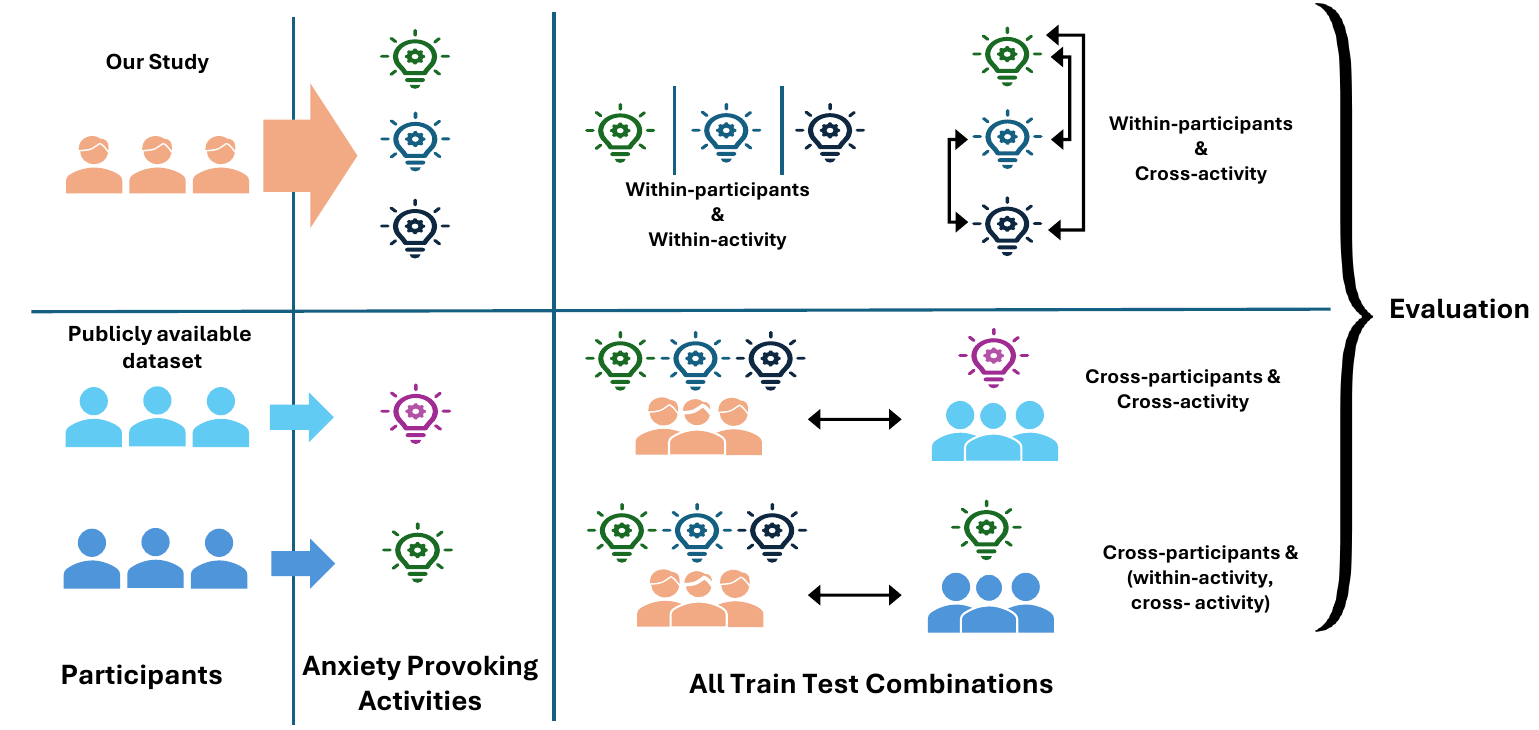}
    \caption{Overview of our contribution: We conducted a study comprising 40 sessions, where each session involved three participants performing three anxiety-inducing activities: a speech activity (A1, shown in green), a group discussion (A2, shown in blue), and an interview (A3, shown in black). Using the ECG and EDA data collected from participants during these sessions, we investigated our research questions on generalizability. Specifically, we examined generalizability within participants \& within activity, as well as within participants \& cross activities. To further explore cross-participant \& cross-activity generalizability and cross-participant \& within activity generalizability, we used two publicly available datasets: WESAD and Anxiety Phase Detection. \textbf{Best viewed in color.}}
    \Description{}
    \label{fig:master_diagram}
\end{figure*}

To address \textbf{RQ2}, we expanded our analysis by incorporating two additional publicly available datasets—WESAD \cite{schmidt2018introducing} and Anxiety Phase Dataset (APD) \cite{senaratne2021multimodal}—in which participants also engaged in anxiety-inducing activities. First, we trained machine learning models on our in-house dataset and tested them on these external datasets. Subsequently, we reversed the process by training models on the WESAD and APD datasets and testing them on our dataset. This approach allowed us to evaluate the transferability and robustness of the models across different datasets.  Focusing specifically on the speech activity present in both our dataset and APD, we found that when models were trained on one dataset and tested on the other, the average best AUROC was 0.62, with a recall of 59.77\% for the anxious class and 60.25\% for the non-anxious class. For other train-test combinations (excluding the common speech activity of our dataset and APD), the average best AUROC was 0.67, with recalls of 62.92\% for the anxious class and 64.69\% for the non-anxious class.
Following are the contributions of our work.
\begin{enumerate}
    \item To our knowledge, this is the first experimental study examining the generalizability of anxiety detection models across different activities (RQ1) and participants groups (RQ2). An indirect benefit of this work is its contribution to the development of just-in-time (JIT) anxiety detection models, designed for real-time prediction of anxiety episodes. Building models that generalize well across varied anxious activities and participant groups is a critical step toward enabling timely, adaptive anxiety interventions, ultimately helping to reduce the treatment gap in mental health care.

    \item We extensively evaluated the generalizability of the anxiety detection models across four distinct scenarios as shown in Figure~\ref{fig:master_diagram}: (i) \textit{Within-participant and within-activity analysis:} Models were trained and tested on the same participants performing the same activity (Section \ref{section: within activity within participants}).
    (ii) \textit{Within-participant and cross-activity analysis:} Models were trained and tested on the same participants but across different activities (Section \ref{section: within participants cross activity}).
    (iii) \textit{Cross-participant and within-activity analysis:} Models were trained and tested on different participants performing the same activity (Section \ref{cross participants}).
    (iv) \textit{Cross-participant and cross-activity analysis:} Models were trained and tested on different participants across varied activities (Section \ref{cross participants}).
    This comprehensive evaluation allowed us to assess the models’ robustness and adaptability across both participants and activities. 

\end{enumerate}

\section{Related Work}

In this section, we begin by underscoring the contributions of UbiComp and HCI (Human-Computer Interaction) to the advancement of mental health research. These interdisciplinary fields have significantly impacted our understanding and treatment of mental health disorders by leveraging technology to collect, analyze, and intervene in real-time. Following this, we discuss recent and pertinent research efforts in the domain of mental health research.

\subsection{Computing in mental health}
Mobile sensing within ubiquitous systems has enabled the passive collection of data through mobile devices like smartphones and wearables in real-world settings, without continuous user attention \cite{rashid2020predicting}. Prior research has leveraged wearables to gather physiological and behavioral data, shedding light on the dynamics of daily life \cite{mishra2020evaluating, mohr2017personal}. In mental health specifically, mobile sensing has been instrumental in exploring physiological and behavioral markers of mental health disorders \cite{mohr2017personal}. These studies have collected a wide range of human physiological signals, such as heart rate \cite{wang2023detecting}, skin conductance \cite{wang2023detecting}, and skin temperature (ST) \cite{shaukat2021detecting}, as well as behavioral signals, like accelerometer \cite{wang2014studentlife, rashid2020predicting}, pedometer \cite{rashid2020predicting}, and GPS data \cite{wang2014studentlife, rashid2020predicting}, to explore associations with mental disorders. For example, to assess anxiety severity, Shaukat-Jali et al. \cite{shaukat2021detecting} used biobehavioral data from wearable sensors, including heart rate (HR), EDA, and ST. Similarly, the pioneering StudentLife \cite{wang2014studentlife} study utilized passively sensed mobile data to analyze mental health and behavioral trends among college students. The mobile sensing studies have employed various wearable devices, including Empatica E4 \cite{disalvo2022reading, wilson2021objective}, Shimmer Sensing \cite{gupta2022total, xu2021cardiacwave}, and Fitbit \cite{doryab2019modeling, lu2018joint}, to collect biobehavioral data, and smartphone applications \cite{rashid2020predicting, wang2014studentlife} to collect user context data for detecting mental health conditions such as stress, anxiety, depression, bipolar disorder, and schizophrenia. Additionally, mobile sensing has been used for context detection, helping to identify whether a participant is in an anxious or neutral state, monitor social dynamics, and assess living environments \cite{wang2023detecting}.

\subsection{Anxiety detection}
%\subsubsection{Summarize all anxiety detection}
Previous research on anxiety detection has utilized diverse data sources, such as social media posts, audio, fMRI (Functional Magnetic Resonance Imaging), EEG (Electroencephalography) signals, etc., for anxiety classification. For example, Kim et al. \cite{kim2020deep} used textual data from Reddit to train an anxiety vs. non-anxiety classifier, achieving an accuracy of 77.81\%. Similarly, Santos et al. \cite{santos2024setembrobr} analyzed Twitter data along with network metrics (e.g., follower count), achieving an F1 score of 0.61. In audio-based studies, Sahu et al. \cite{sahu2024unveiling} and Salekin et al. \cite{salekin2018weakly} used impromptu speech audio data to detect anxiety, achieving accuracies of 85.71\% and 90\%, respectively. Matteo et al. \cite{di2021smartphone} examined smartphone-recorded environmental audio and found a positive correlation between vision-related words and social anxiety. In physiological approaches, Ezzi et al. \cite{al2021severity} used EEG signals collected during a social performance task to classify the severity of social anxiety disorder (SAD), with a peak accuracy of 92.86\%. In terms of fMRI-based approaches, Liu et al. \cite{liu2015multivariate} used fMRI to classify SAD versus healthy controls, achieving an accuracy of 82.5\%, while Doehrmann et al. \cite{doehrmann2013predicting} employed fMRI to predict treatment responses in SAD participants.

%\subsubsection{Summarize only wearable sensors anxiety detection work}
%Compared to the previously discussed data sources, the last decade has seen a surge in mental health assessment studies leveraging passively collected data (e.g., wearable sensors).

Over the past decade, there has been a surge in mental health assessment studies that leverage passively collected data from wearable sensors. 
Wearables collected data such as electrocardiogram (ECG), PPG, EDA, ST, etc., have emerged as an important data source for anxiety screening, where machine learning models are used to learn patterns like heart rate variability, skin conductance, etc., in anxious and non-anxious participants. In particular, studies on anxiety detection using wearables have concentrated on (i) distinguishing individuals with and without anxiety disorders, (ii) predicting anxiety severity, and (iii) classifying anxiety states, such as baseline/rest versus stress period (e.g., TSST period). Following are brief summaries of several such studies. 

Coutts et al. \cite{coutts2020deep} collected four weeks of ECG data from 652 participants and computed time-domain, frequency-domain, and standard Heart Rate Variability (HRV) measures. Using HRV measures alongside weekly self-reported anxiety levels, they trained a deep learning model to predict high and low anxiety during both day and night, achieving accuracies of 63.2\% for daytime and 69.4\% for nighttime predictions. Another study (Ihmig et al. \cite{ihmig2020line}) used ECG, EDA, and respiration data collected from 80 spider-fearful participants aged 18 to 40 as they watched 16 video clips featuring spiders. By training machine learning models on features derived from ECG, EDA, and respiration, they achieved an accuracy of 89.8\% in distinguishing high vs. low anxiety (two-level classification) and 74.4\% in distinguishing low, medium, and high anxiety (three-level classification). Similarly, Šalkevičius et al. \cite{vsalkevicius2019anxiety} asked 30 participants to deliver a five-minute speech in front of a virtual audience while passively collecting biobehavioral data (blood volume pressure, EDA, and ST). Afterward, participants provided self-reported anxiety scores using the Subjective Units of Distress Scale (SUDS) as a ground truth measure. Using 10-fold cross-validation, they achieved an accuracy of 86.3\% in identifying four levels of anxiety.

Schmidt et al. \cite{schmidt2018introducing} designed a study involving 15 participants who went through six distinct phases: baseline, amusement, two meditation sessions, stress, and rest. The stress-inducing task, the Trier Social Stress Test (TSST), required participants to deliver a five-minute impromptu speech followed by five minutes of reverse counting from 2023 to zero, with participants restarting if they made an error. ECG, EDA, electromyography (EMG), skin temperature (ST), blood volume pulse (BVP), and accelerometer (ACC) data were collected throughout. Machine learning models used to classify these states achieved an accuracy of 80\% in three-class classification (baseline vs. stress vs. amusement) and 93\% in two-class classification (stress vs. non-stress). Bhatti et al. \cite{bhatti2024attx} later utilized Schmidt's \cite{schmidt2018introducing} dataset and applied deep learning models, enhancing accuracy to 89.57\% for three-class classification and 93.70\% for two-class classification.

In line with existing work on anxiety detection, we collected a unique biobehavioral dataset corresponding to three anxiety-inducing activities using wearables in controlled settings. The study evaluates the performance of various machine learning methods for classifying anxious and non-anxious participants. Our findings will contribute to and enhance current mental health research. 

\subsection{Generalizability in mental health}
Although machine learning models trained on biobehavioral data have emerged as important tools for predicting mental disorders, their performance remains uncertain when tested on new participant groups.
Generalizability in mobile sensing models, particularly for mental health and behavioral monitoring, faces significant challenges due to demographic, cultural, and behavioral diversity. Xu et al. \cite{xu2023globem} demonstrated that depression detection models trained on data from one cohort (collected in Year 1) underperformed when tested on data from the same institution in Year 2 (cross-year testing). They also observed that models trained on data from one institution did not generalize well when applied to another institution’s data. The authors attributed this to behavioral differences both within and between populations. However, they found that performance improved when there were overlapping participants across the years. Similarly, Kammoun et al. \cite{kammoun2023understanding} found that models predicting social contexts for eating behaviors benefited from diversity-aware designs, as differences in lifestyle, demographic factors, and culture influenced model transferability and introduced biases. 

In another mobile sensing work on speech-based mental health models, Pillai et al. \cite{pillai2024investigating} examined the limitations of machine learning and deep learning models for detecting suicidal ideation across varied datasets. They found that traditional models struggled under distribution shifts, particularly when trained on single-source datasets. To address this, they proposed a sinusoidal similarity sub-sampling (S3) method, which optimizes cross-dataset performance without requiring extensive labels from the target data. Similarly, Meegahapola et al. \cite{meegahapola2023generalization} studied the generalizability of mood inference models on a dataset from 678 participants across eight countries, where an Android application collected passive mobile sensor data and self-reported questionnaires. They found that models trained and tested on data either from the same or different country performed poorly.

These studies underscore the challenges of generalizability in mobile sensing applications for mental health, where cultural, lifestyle, and data distribution shifts introduce substantial complexity. In this study, we evaluate, for the first time to our knowledge, the generalizability of anxiety detection models using mobile-sensed biobehavioral data.

\section{Study Design and Data Collection}
To address our research question on the generalizability of the anxiety detection model, we designed a study in which participants underwent three anxiety-provoking activities. The study was developed by a team of computer scientists, clinical psychologists, and clinical psychiatrists. All study procedures were approved by the Institutional Review Board. The study was conducted by a research assistant (certified good clinical practitioner, PhD student) under the guidance of a clinical psychiatrist. 

\subsection{Participants}
We recruited $111$ undergraduate and postgraduate students from our home institute for this study. Inclusion criteria required participants to be 18 years or older and proficient in English. Recruitment was conducted by emailing the student community, which included a brief introduction about the study and the link to a form for those interested in participating. Subsequently, the research assistant (RA) confirmed each participant’s involvement through email and text messages. Nine participant's data was incomplete, so we dropped it during the analysis. Table \ref{tab:demographic_information} presents the demographic information of 102 study participants, which includes 70 males and 32 females, with a mean age of 21.00.

% \begin{figure}[!t]
%     \centering
%     \begin{minipage}{0.48\textwidth} % Adjust width as needed
%         \centering
%         \captionof{table}{Participants demographic information. 
%         Acronyms M, F, U, and R stand for male, female, urban, and rural, respectively.}
%         \begin{tabular}{cccc} \toprule
%              Participants (\#) & Age & Gender & Home location\\
%              & ($\mu$, $\sigma$) & (\# M, \# F) & (\# U, \# R) \\ \midrule
%              102 & (21.0, 2.68) & (70, 32) & (78, 24) \\ \bottomrule
%         \end{tabular}
%         \label{tab:demographic_information}
%     \end{minipage}
%     \hfill
%     \begin{minipage}{0.48\textwidth} % Adjust width as needed
%         \centering
%         \includegraphics[width=0.9\linewidth]{images/Diagrams/sensor_placement.pdf}
%         \caption{Position of ECG and EDA electrodes of shimmer sensors on participants' chest and fingers, respectively.}
%         \label{fig:sensor_placement}
%     \end{minipage}
% \end{figure}

\begin{table}
    \centering
        \caption{Participants demographic information.
        Acronyms M, F, U, and R stand for male, female, urban, and rural, respectively. 
        }
    \begin{tabular}{cccc} \toprule
         Participants (\#) &  Age &Gender&  Home location\\
 &($\mu$, $\sigma$) & (  \#  M,   \#  F)& (  \# U,  \# R)\\ \midrule
         102&  (21.0, 2.68)&  (70, 32)&  (78, 24)\\ \bottomrule
    \end{tabular}
    \label{tab:demographic_information}
\end{table}
\subsection{Sensor deployment}
In this study, we used Shimmer\footnote{\url{https://shimmersensing.com/}} ECG\footnote{\url{https://shimmersensing.com/product/shimmer3-ecg-unit-2/}} and EDA\footnote{\url{https://shimmersensing.com/product/shimmer3-gsr-unit/}} sensors, clinically validated devices commonly used in similar research settings for data collection. The ECG sensors were placed around the participant's chest, while the EDA sensors were placed on the non-dominant wrist. Sensor placement was conducted by trained researchers involved in the data collection procedures. Before applying the electrodes, the designated areas were cleansed with an alcohol swab to ensure optimal adhesion and signal quality.

\begin{figure}
    \centering
    \includegraphics[width=0.5\linewidth]{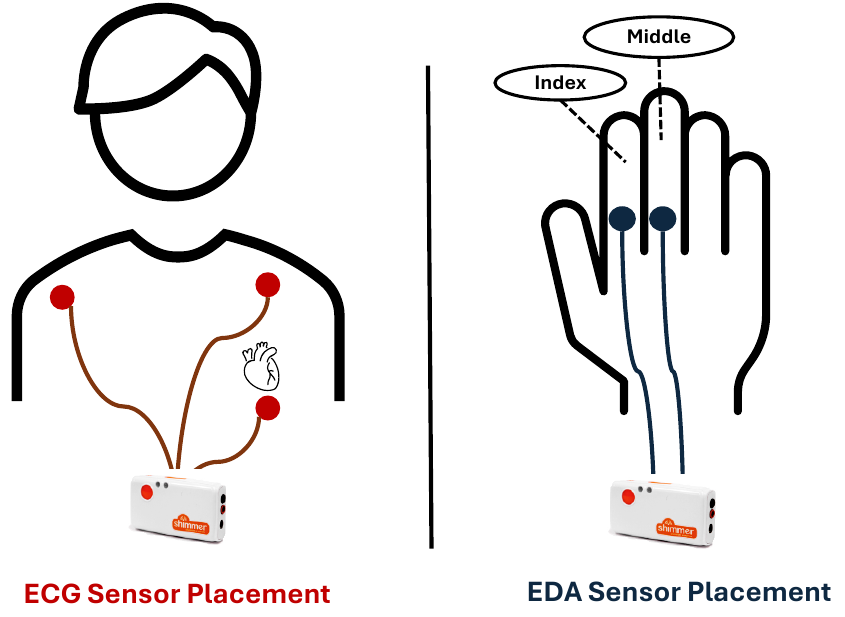}
    \caption{Position of ECG and EDA electrodes of shimmer sensors on participants' chest and fingers, respectively.}
    \Description{}
    \label{fig:sensor_placement}
\end{figure}

For ECG measurements, we used 3-pin electrodes. Two electrodes were positioned below the shoulders on the right and left sides, and a third was placed below the heart (see Figure \ref{fig:sensor_placement}). EDA electrodes were attached to the index and middle fingers. The RAs ensured that ECG and EDA electrodes were firmly attached to minimize movement artifacts. The sensors were configured to collect ECG and EDA data at a sampling rate of 1024 Hz.

\textit{}
\subsection{Study procedure}
We invited three participants to each study session, with each session facilitated by 3-4 members of our research team in a dedicated room. A lead RA conducted the study, assisted by two RAs responsible for sensor placement. Prior to inviting participants, the RAs ensured that the participants were not acquainted with one another. The participants were contacted one day before the study and again one hour prior to their session to confirm their availability in the given time slot and the study location. 

\begin{figure*}[!h]
  \centering
  \includegraphics[width=\textwidth, height=0.2\textheight, keepaspectratio]{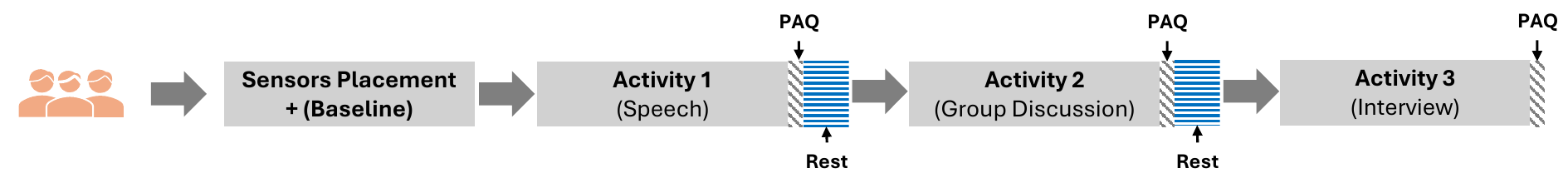}
  \caption{ 
Activity sequence followed in each study session. PAQ pointing to shaded blocks (grey color) \includegraphics[width=0.20cm, height=0.35cm]{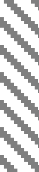} denote the instances at which participants filled the post activity questionnaire (PAQ). Shaded blocks (blue color) \includegraphics[width=0.20cm, height=0.35cm]{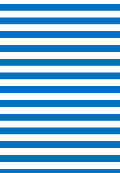}, represent the rest period before starting the next activity.}
\Description{}
  \label{fig:study_flow}
\end{figure*}

Upon arrival, participants were seated at designated spots around a rectangular table with rounded edges. The lead RA provided a brief overview of the study and obtained informed consent from each participant. After consent was secured, the sensors were placed. After the sensor placement, there were three participants and the lead RA present in the room. The participants were labeled as P1, P2, and P3, with P1 seated on the left side of the RA, P2 opposite the RA, and P3 on the right side. Thus, two participants faced each other while the third faced the RA. Figure~\ref{fig:study_flow} shows the sequence of steps followed in each study session.

Next, the RA initiated a baseline period, during which participants were asked to sit quietly for two minutes. The baseline period aimed to help participants acclimate to the lab setting and minimize any pre-existing physiological changes that might have arisen from the study briefing and sensor placement. After completing the baseline, participants were informed about the \textbf{Activity \# 1}: an impromptu speech lasting 2.5 minutes. P1 was given a speech topic with 30 seconds for preparation. After this period, the RA prompted P1 to begin the speech. Upon completion, P1 filled out the post-activity questionnaire (PAQ), discussed below. A similar procedure was followed for P2 and P3, each with different speech topics.
After the speech activity, participants were informed about the \textbf{Activity \# 2}: a group discussion (GD) where all three were asked to express their opinions on a given topic. The GD topic was shared with all participants, along with 30 seconds of preparation time, and they were informed that the discussion would last for six minutes. After the preparation, the RA randomly selected one participant to start the discussion. Following the six-minute GD, participants were asked to complete the PAQ.

Next, P2 and P3 were asked to step outside the room while P1 was briefed about the \textbf{Activity \# 3}: a three-minute interview. This interview was conducted by a senior RA dedicated solely to this task. After the interview, P1 filled out the PAQ and then stepped outside. The process was repeated for P2 and then P3. Once all interviews were complete, the participants were thanked for their participation. Finally, the RA removed the sensors from each participant and provided refreshments as compensation for their involvement in the study. The sensor data was transferred to a lab computer after each session. 

\begin{table*}[]
\small
\caption{Post-activity questionnaire (PAQ) collected after the Activity 1, Activity 2, and Activity 3.}
\label{tab:paq}
\begin{tabular}{lc}
\toprule
\textbf{Questions}                                                                                         & \textbf{Response (Scale of 1 to 5)}   \\ \midrule
Q1. I felt extreme nervousness in the last activity.& Strongly Disagree (1) ... Strongly Agree (5) \\
Q2. During the last activity, how did you feel about yourself?& Unhappy/Negative (1) ... Happy/Positive (5)  \\
Q3. I wish I could have avoided the last activity as it produced      anxiety in me.& Strongly Disagree (1) ... Strongly Agree (5) \\
Q4. During the last activity, I was worried about being scrutinized by others.& Strongly Disagree (1) ... Strongly Agree (5) \\
Q5. During the last activity, I thought my behavior was negatively evaluated.& Strongly Disagree (1) ... Strongly Agree (5) \\ \bottomrule
\multicolumn{2}{l}{\textbf{Note}: \textit{The responses for Q2 were reversed during analysis.}}
\end{tabular}
% \caption*{\footnotesize \raggedright \textbf{Note}: \textit{The responses for Q2 were reversed during analysis.}}
\end{table*}

\subsection{Post activity questionnaire (PAQ)}

To assess participants' anxiety during each activity, we designed a PAQ consisting of five questions. Validation of the PAQ was conducted in two stages: first, an initial version was reviewed by nine clinical psychiatrists, independent of our study, whose feedback helped refine the questionnaire. In the second stage, this revised version was further validated by 11 additional clinical psychiatrists. Additionally, we piloted the questionnaire with 40 undergraduate and postgraduate students to ensure the clarity of questions, response options, and wording.

Table \ref{tab:paq} shows the finalized PAQ used in this study. The final PAQ helped us in assessing the participant's levels of experienced anxiety, emotional arousal, activity avoidance, excessive worry, and perceived negative evaluation of their performance. Participants completed the PAQ after each activity to report their anxiety levels experienced during the activities. The cumulative PAQ score represents each participant’s self-reported anxiety level, which we used to classify participants into anxious and non-anxious groups in each activity (discussed in detail in Section \ref{section: Ground Truth}).

\section{Methodology}
\subsection{Data cleaning and feature extraction} \label{section: feature_extraction}
After collecting the data, we inspected the ECG and EDA signals to assess their quality. During this process, we noticed the presence of noise, likely caused by body movements, which further required filtering. Filtering involves removing unwanted noise from the signals. We reviewed relevant literature \cite{wang2023detecting,urrestilla2020measuring,shaffer2017overview,reyero2022heart} to choose appropriate filtering parameters and examined the quality of the filtered signals. Empirically, we found that a bandpass filter with a range of 1-49 Hz worked effectively for all three activities in our ECG dataset. Similarly, a low-pass filter with a 5 Hz cutoff for the EDA signals produced the best results. Next, we segmented the filtered ECG and EDA signals using sliding windows of 60 seconds with a 0.25-second shift, following the approach of Schmidt et al. \cite{schmidt2018introducing}. Feature extraction was then performed on each window as described below.

% We applied a peak detection algorithm to the filtered ECG signals. 
\textit{ECG features:} We explored various peak detection algorithms available in the Neurokit\footnote{\url{https://github.com/neuropsychology/NeuroKit}} and Biosppy\footnote{\url{https://github.com/scientisst/BioSPPy}} Python packages. After visually inspecting the peaks identified by different algorithms, we found that the Hamilton segmented \cite{hamilton2002open} performed the best. However, in some cases, even after filtering, the ECG signals remained slightly noisy, affecting peak detection accuracy. As a result, we decided to exclude the data from these participants, regardless of the activity. After detecting the peaks, we used the Neurokit package to extract heart rate variability (HRV) metrics \cite{makowski2021neurokit2}. The HRV features were calculated in both time and frequency domains. Additionally, we also computed non-linear indices of HRV and Respiratory Sinus Arrhythmia (RSA).

\textit{EDA features:} The filtered windowed EDA signals were decomposed into phasic components, also known as Skin Conductance Response (SCR), and tonic components, also known as Skin Conductance Level (SCL). The phasic component represents faster changes, while the tonic component reflects slower alterations in skin conductance. For feature extraction, we followed the methods outlined by Varun et al. \cite{gashi2020detection} and Wang et al.\cite{wang2023detecting}, computing statistical features such as the mean, standard deviation, minimum, and maximum of SCR (duration, amplitude, rise time, and recovery time), raw signal, cleaned signal, and SCL. Additionally, we calculated the median, range, and slope of SCR, the mean and standard deviation of the first and second derivatives of SCR, the area under the curve of SCR, and the mean, median, and standard deviation of wavelet coefficients at 4 Hz, 2 Hz, and 1 Hz.

\begin{table*}[t]
\centering
\small
\caption{List of the final set of features used for analysis. Throughout the paper, we refer to the feature sets with their abbreviations, F1, F2, F3, F4, and F5. Features F1 to F4 were derived from the ECG signal as explained in  \cite{makowski2021neurokit2, wang2023detecting}, while F5 is derived from the EDA signal as explained in \cite{gashi2020detection, wang2023detecting}.}
\label{tab:feature}
\begin{tabular}{@{}p{4cm}lp{10cm}@{}}
\toprule
\textbf{Feature set}        &\#& \textbf{Features}                                         \\ \midrule
Time Domain HRV \textbf{(F1)}       &5& MeanNN, SDNN, MadNN, MinNN, TINN                 \\
Frequency Domain HRV \textbf{(F2)}  &4& LF, HF, VHF, LF-HF                                \\
Non Linear Indices HRV \textbf{(F3)}  &21&
  SD1, CSI\_Modified, PIP, PAS, GI, PI, C1d, C2d, MFDFA\_alpha1\_Width, MFDFA\_alpha1\_Peak, MFDFA\_alpha1\_Mean, MFDFA\_alpha1\_Max, MFDFA\_alpha1\_Fluctuation, ApEn, SampEn, ShanEn, MSEn, CMSEn, HFD, KFD, LZC \\
RSA HRV   \textbf{(F4)}             &6& RecurrenceRate, DiagRec, Determinism, L, W, WMax \\
% EDA \textbf{(F5)}  &15&
%   mean(SCR - duration, amplitude, RiseTime, first derivative, second derivative, wavelet 4Hz), Max(SCR - duration, amplitude, RiseTime, RecoveryTime), Min(SCR duration, amplitude), Median (SCR - values, wavelet 4Hz), SCR wavelet 4Hz standard deviation, SCR slope \\ \bottomrule
EDA \textbf{(F5)}  &15&
  mean SCR duration, mean SCR amplitude, mean SCR RiseTime, mean SCR first derivative, mean SCR second derivative, mean SCR wavelet 4Hz, maximum SCR duration, maximum SCR amplitude, maximum SCR RiseTime, maximum SCR RecoveryTime, minimum SCR duration, minimum SCR amplitude, median SCR values, median SCR wavelet 4Hz, SCR wavelet 4Hz standard deviation, SCR slope \\ \bottomrule
\end{tabular}
\end{table*}

\subsection{Feature selection}
After feature extraction, the number of features was as follows: HRV – 25 time domain features, 10 frequency domain features, 56 non-linear indices, and 15 RSA features; EDA had 46 features. However, we found some features with missing values, which we removed. Next, we applied a correlation analysis to eliminate highly correlated features, using a correlation threshold of 0.75 as suggested in  \cite{rashid2020predicting}. The final set of features used in this work is shown in Table \ref{tab:feature}. Throughout the paper, we refer to the five feature sets of Time Domain HRV, Frequency Domain HRV, Non-Linear Indices HRV, RSA HRV, and  EDA as \textbf{F1}, \textbf{F2}, \textbf{F3}, \textbf{F4}, and \textbf{F5}, respectively.

%After feature extraction, the number of features was as follows: HRV – 25 time domain features, 10 frequency domain features, 56 non-linear indices, and 15 RSA features; EDA had 46 features. However, we found some features with missing values, which we removed. Next, we applied a correlation analysis to eliminate highly correlated features, using a correlation threshold of 0.75\hl{citeme}. The final set of features used in this work is shown in Table \ref{tab:feature}. 

\section{Data Characteristics}
In this section, we used statistical analysis, i.e., descriptive and inferential statistics, to understand the self-reported PAQ and Shimmer sensor (ECG, EDA) data. 

% \vspace{-10pt}

\subsection{PAQ data distribution}
\begin{figure*}[!t]
    \centering
    % First Row:
    \begin{minipage}[b]{0.48\textwidth}
        \centering
        \subcaptionbox{Activity 1 \label{fig:PAQ_F1}}{
            \includegraphics[width=0.9\linewidth]{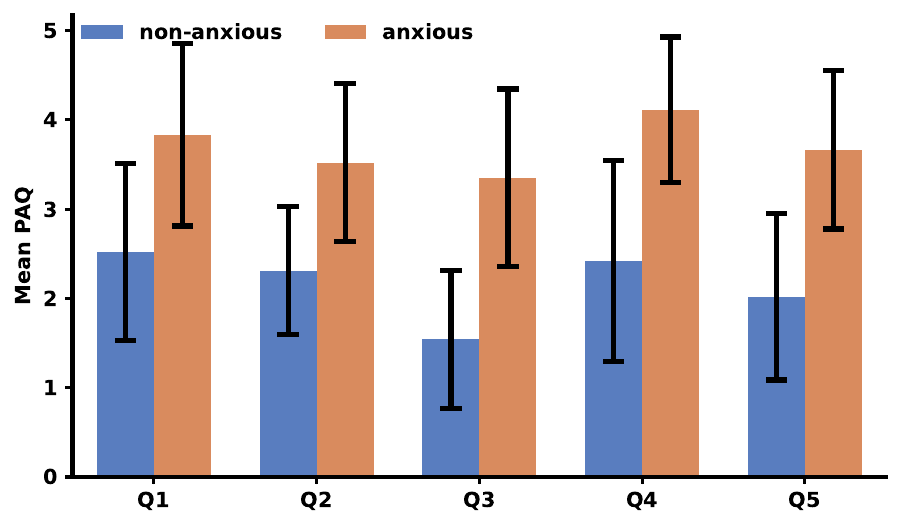}
        }
    \end{minipage}
    \begin{minipage}[b]{0.48\textwidth}
        \centering
        \subcaptionbox{Activity 2 \label{fig:PAQ_F2}}{
            \includegraphics[width=0.9\linewidth]{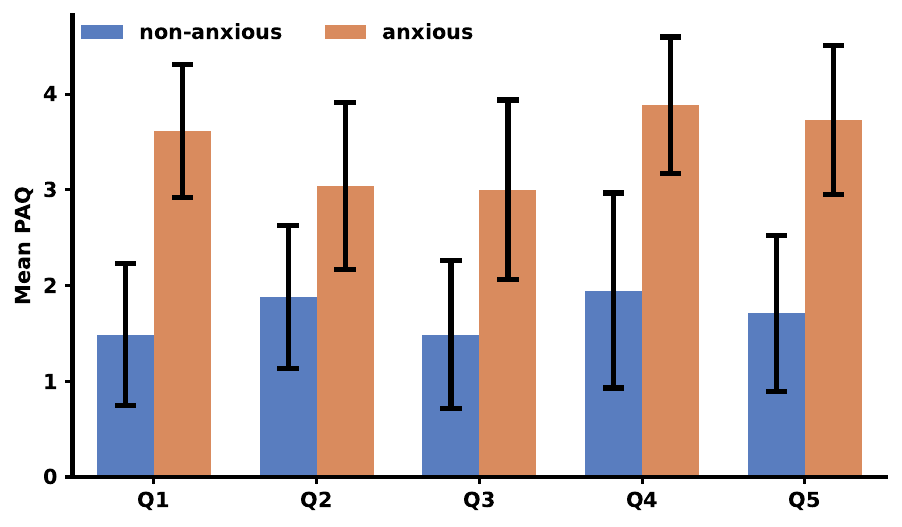}
        }
    \end{minipage}
    
    \vspace{0.5cm} % Space between rows

    \begin{minipage}[b]{0.48\textwidth}
        \centering
        \subcaptionbox{Activity 3 \label{fig:PAQ_F3}}{
            \includegraphics[width=0.9\linewidth]{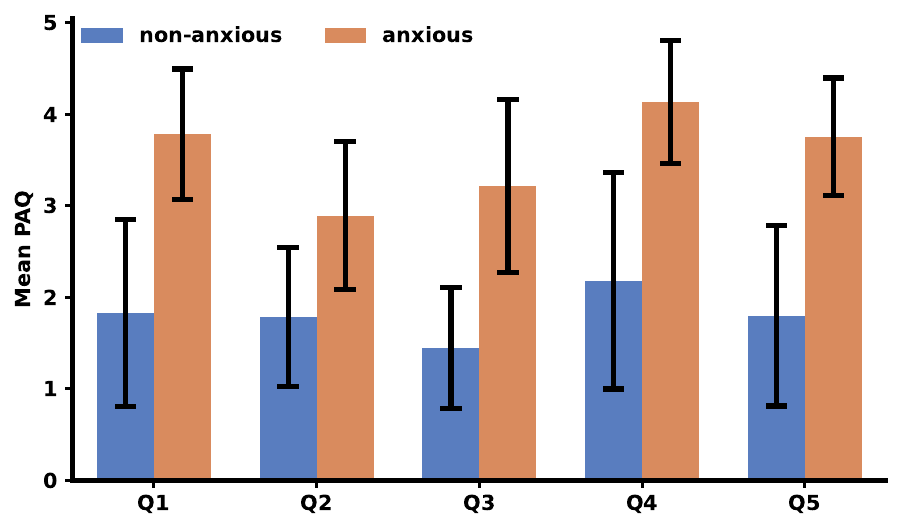} 
        }
    \end{minipage}
    \begin{minipage}[b]{0.48\textwidth}
        \centering
        \subcaptionbox{For all 3 activities\label{fig:PAQ_F4}}{
            \includegraphics[width=0.9\linewidth]{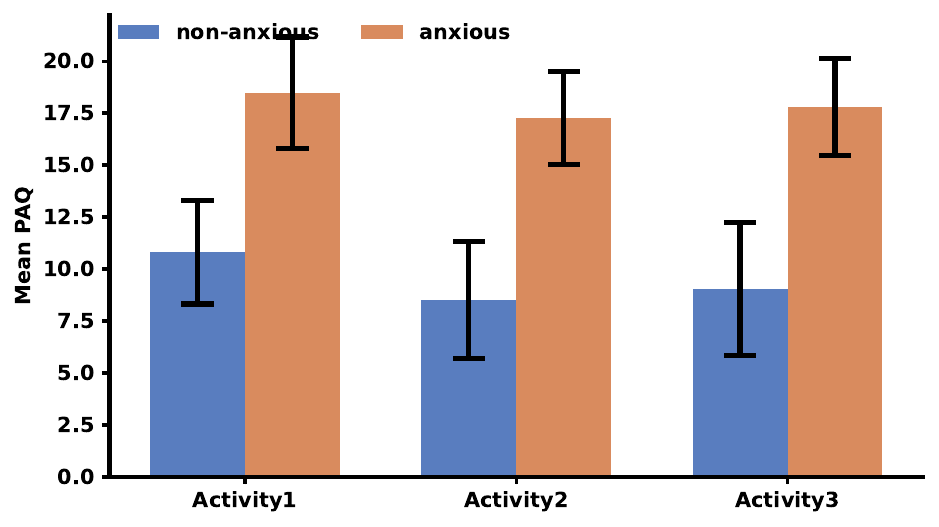} 
        }
    \end{minipage}
    
    \caption{Mean PAQ scores of anxious and non-anxious participants for five different questions (i.e., Q1, Q2, Q3, Q4, and Q5) during \textbf{(a)} Activity 1 (Speech), \textbf{(b)} Activity 2 (Group discussion), and \textbf{(c)} Activity 3 (Interview). \textbf{(d)} shows the mean of all 5 questions for different activity 1, 2, and 3, separately.}
    \Description{}
    \label{fig:sr_barplot}
\end{figure*}

Figures \ref{fig:sr_barplot}(a), (b), and (c) show the distribution of the mean self-reported PAQ scores for different questions during activities 1, 2, and 3, respectively, for both the anxious and non-anxious groups. Overall, the anxious group reported higher levels of anxiety, negativity, likelihood of activity avoidance, worry about being scrutinized, and fear of negative evaluation. Particularly, in Activity 1 (i.e., Speech), the non-anxious group reported the highest mean self-reported score for Q1, while the anxious group reported the highest score for Q4, suggesting that anxious participants were most concerned about being scrutinized during the speech performance. During Activities 2 (i.e., Group discussion) and 3 (i.e., interview), both groups reported the highest concern about being scrutinized (i.e., Q4). 
Furthermore, Figure \ref{fig:sr_barplot}(d) illustrates the average of all five question scores for each activity, comparing anxious and non-anxious groups. The Figure shows that the speech activity triggered the highest anxiety levels, while the group discussion elicited the lowest. Table \ref{tab:cohen_for_activities} presents the T-statistics (TS) and Cohen's d (CD) for comparing the anxious and non-anxious groups' self-report scores across all activities. The highest (3.51) Cohen's d was observed for Activity 2, compared to Activities 1 and 3, indicating a greater difference in self-reported PAQ scores between the anxious and non-anxious groups during this activity.

\begin{table}[]
    \centering
\caption{T-statistic (TS) and Cohen’s-d (CD)  computed between anxious and non-anxious groups based on cumulative PAQ.}
\label{tab:cohen_for_activities}
    \begin{tabular}{cccc} \toprule
         &  \textbf{t-statistic (TS)} &  \textbf{Cohen’s-d (CD)} & \textbf{p-value} \\ \midrule
         \textit{Activity 1}&  15.9&  3.01& <0.05\\ 
         \textit{Activity 2}&  15&  3.51& <0.05\\ 
         \textit{Activity 3}&  14.6&  3.09& <0.05\\ \bottomrule
    \end{tabular}
\end{table}

\subsection{Sensor data distribution} \label{section: data_similarity}
Table \ref{tab:sensors_stats} presents the TS and CD values for the sensor-extracted features, comparing the anxious and non-anxious groups across each activity. The table highlights the top three features with the highest Cohen’s d for each activity. Most features showed either a low or medium effect size. In Activities 2 and 3, MeanNN had a higher effect size, indicating lower MeanNN values in anxious individuals. Although the same trend was observed for Activity 1, the difference was not significant enough to be considered. This finding aligns with prior studies that have reported lower HRV values in anxious individuals compared to non-anxious groups \cite{chalmers2014anxiety}.

\begin{figure*}[!t]
    \centering
    % First Row: F1, F2, F3
    \begin{minipage}[b]{0.32\textwidth}
        \centering
        \subcaptionbox{F1 \label{fig:F1}}{
            \includegraphics[width=\linewidth]{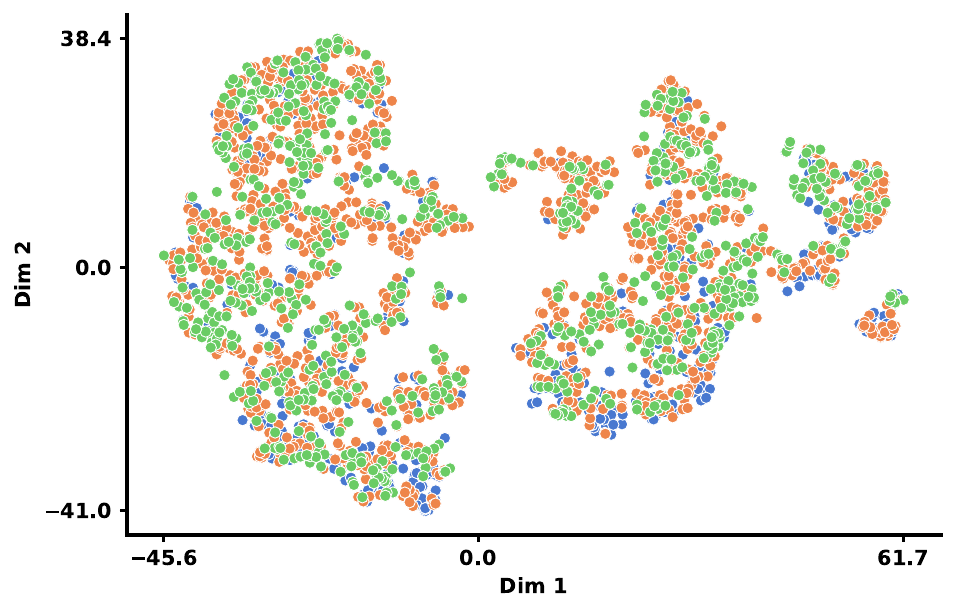}
        }
    \end{minipage}
    \begin{minipage}[b]{0.32\textwidth}
        \centering
        \subcaptionbox{F2 \label{fig:F2}}{
            \includegraphics[width=\linewidth]{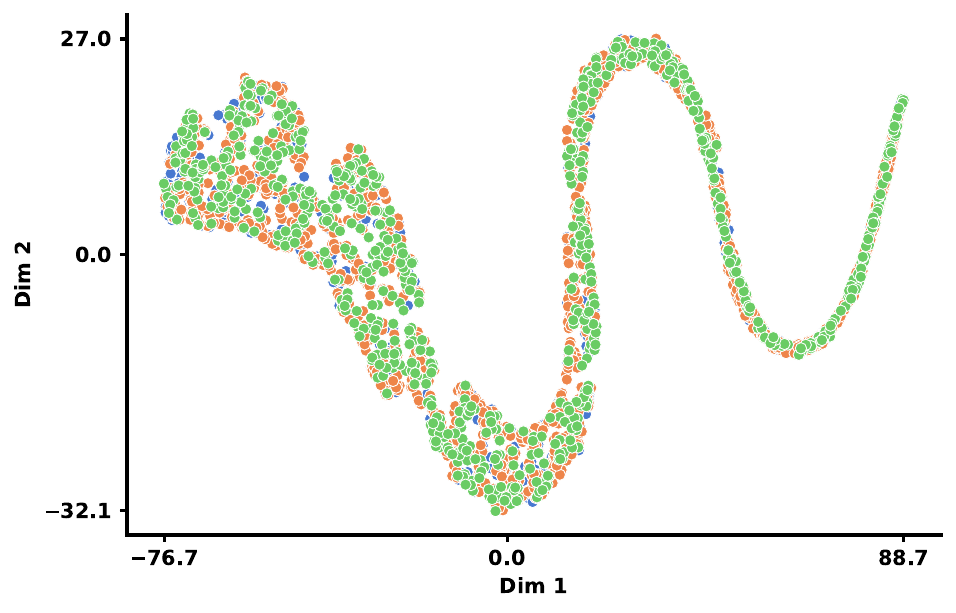}
        }
    \end{minipage}
    \begin{minipage}[b]{0.32\textwidth}
        \centering
        \subcaptionbox{ F3 \label{fig:F3}}{
            \includegraphics[width=\linewidth]{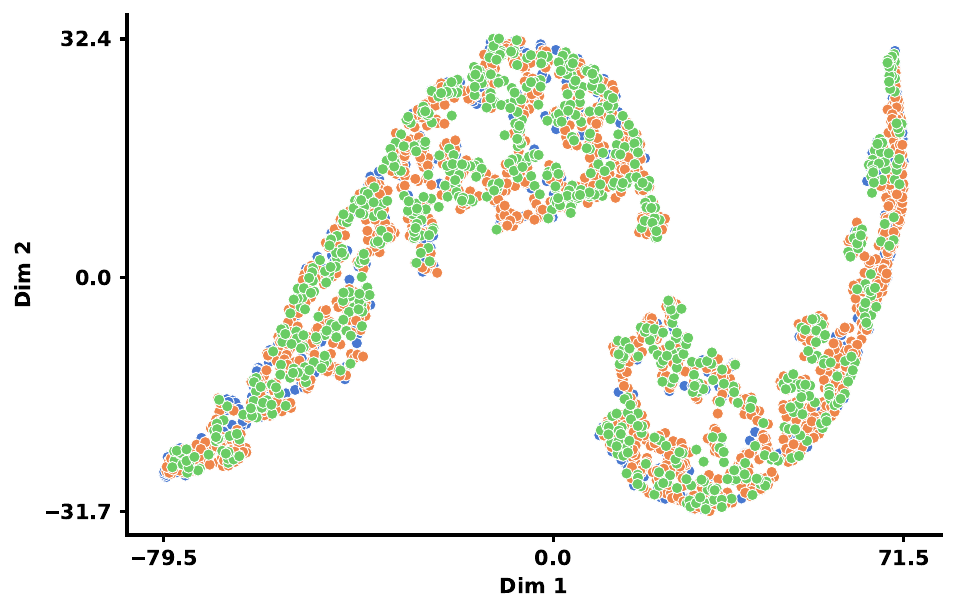}
        }
    \end{minipage}
    
    \vspace{0.5cm} % Space between rows

    % Second Row: F4, F5, Legend
    \begin{minipage}[b]{0.32\textwidth}
        \centering
        \subcaptionbox{F4 \label{fig:F4}}{
            \includegraphics[width=\linewidth]{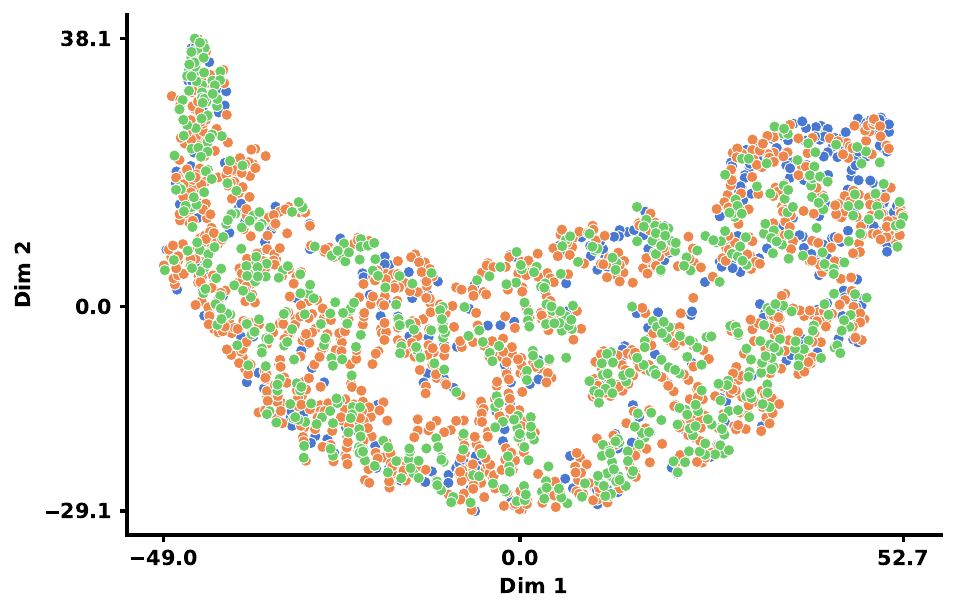} 
        }
    \end{minipage}
    \begin{minipage}[b]{0.32\textwidth}
        \centering
        \subcaptionbox{F5 \label{fig:F5}}{
            \includegraphics[width=\linewidth]{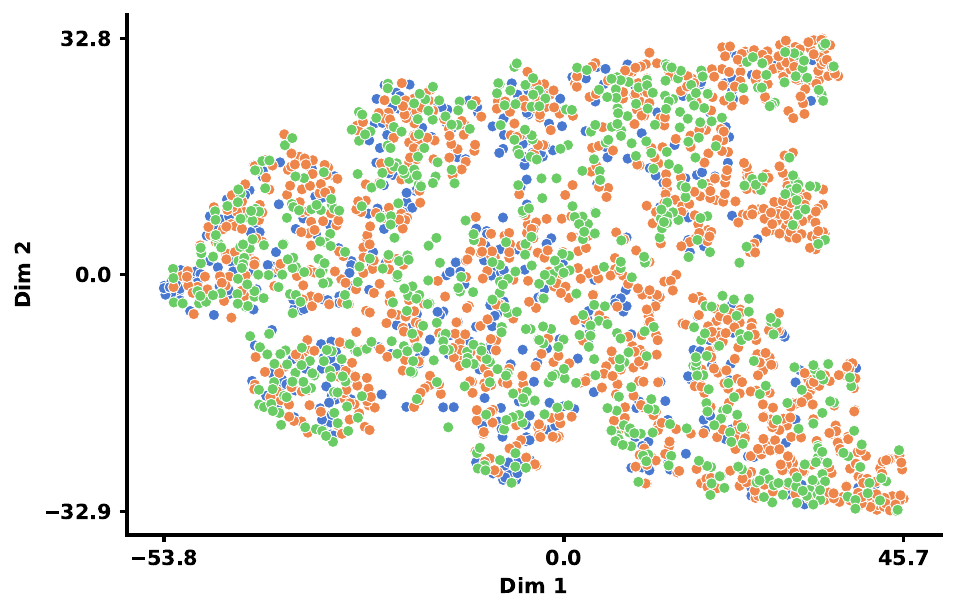} 
        }
    \end{minipage}
    \begin{minipage}[b]{0.32\textwidth}
        \centering
            \includegraphics[width=\linewidth]{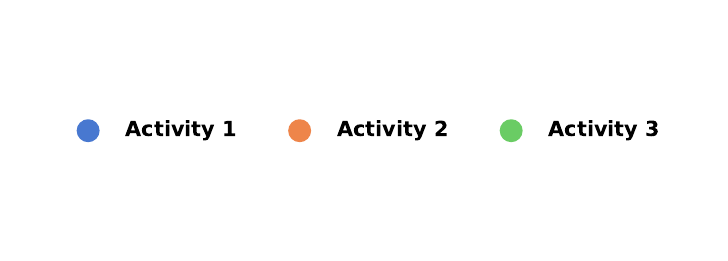} 
            \phantomsubcaption
    \end{minipage}

    \caption{t-SNE plots, i.e., low-dimensional representation of high-dimensional features sets \textbf{(a)} F1, \textbf{(b)} F2, \textbf{(c)} F3, \textbf{(d)} F4, and  \textbf{(e)} F5 collected during Activity 1 (blue colour), Activity 2 (Orange color),  and Activity 3 (green color). \textbf{Best viewed in colour.}}
    \Description{}
    \label{fig:t-SNE-feature-sets}
\end{figure*}

Since our objective was to evaluate the generalizability of the anxiety detection models, we examined the characteristics of our data visually using t-distributed Stochastic Neighbor Embedding (t-SNE) plots \cite{van2008visualizing}. t-SNE is a well-known dimensionality reduction technique used to visualize high-dimensional data in lower dimensions. Figure \ref{fig:t-SNE-feature-sets} shows the two dimensional representation plots for feature sets F1, F2, F3, F4, and F5 for activities 1, 2, and 3. The plots reveal that all five feature sets from activities 1, 2, and 3 exhibit similar characteristics and are comparable in nature.  

\begin{figure}
    \centering
    \includegraphics[width=0.5\linewidth]{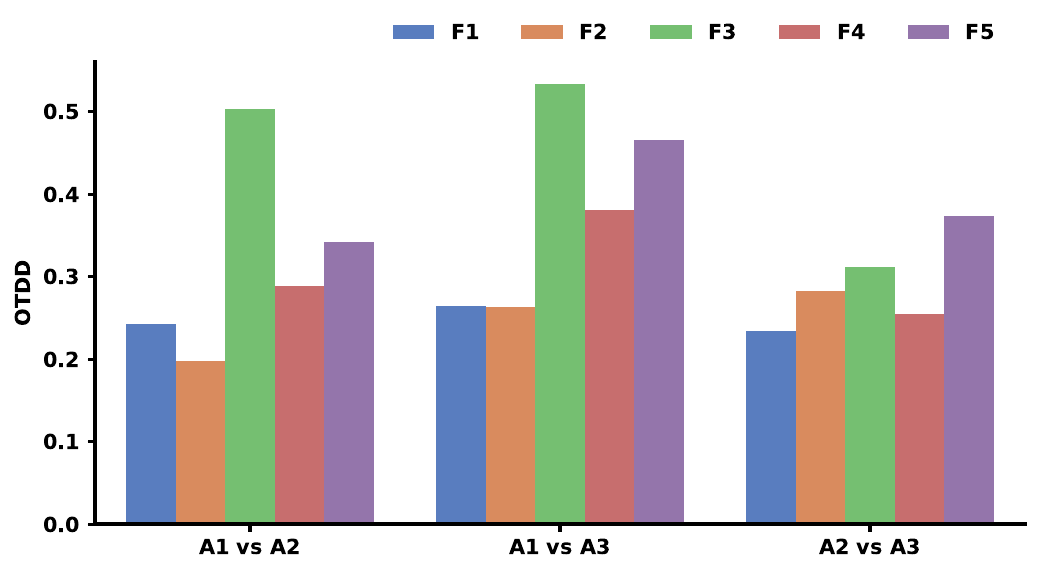}
    \caption{Optimal Transport Dataset Distance (OTDD) values computed between different pairs of datasets (i.e., activities in our case) on five different feature sets (F1, F2, F3, F4, and F5). \textbf{Best viewed in colour.}}
    \Description{}
    \label{fig:otdd-plots}
\end{figure}

To further validate the similarity of the data, we also computed the Optimal Transport Dataset Distance (OTDD\footnote{\url{https://www.microsoft.com/en-us/research/blog/measuring-dataset-similarity-using-optimal-transport/}}) \cite{alvarez2020geometric}. It is based on optimal transport theory and is used to calculate the distance between two datasets. The range of OTDD is generally $[0, \infty)$, where values closer to 0 indicate higher similarity, and it increases with dissimilarity. Figure \ref{fig:otdd-plots} presents the OTDD values computed for feature sets \textit{F1, F2, F3, F4,} and \textit{F5} across different dataset (activity) combinations (A1 versus A2, A1 versus A3, and A2 versus A3). Across all dataset/activities combinations, we observed the lowest OTDD values for feature sets \textit{F1} and \textit{F2}, indicating higher similarity in these sets.

\noindent{\textbf{\textit{Takeaway}}}: The t-SNE plots and the OTDD values indicate that the datasets (A1, A2, and A3) are similar and share the same data characteristics. This suggests that prediction models should be generalizable, meaning that a model trained on one dataset can be effectively tested on another.

\begin{table}[t]
\centering
\caption{T-statistics (TS) and Cohen’s d (CD) for different features during Activity 1, Activity 2, and Activity 3. * represents results significant at p<0.05}
\label{tab:sensors_stats}
\resizebox{\columnwidth}{!}{%
\begin{tabular}{@{}lccclccclcc@{}}
\toprule
\multicolumn{4}{c}{\textbf{Activity 1}}                   & \multicolumn{4}{c}{\textbf{Activity 2}}                    & \multicolumn{1}{c}{\textbf{Activity 3}}  & \multicolumn{1}{l}{} \\ \cmidrule{1-3} \cmidrule{5-7} \cmidrule{9-11}
\textit{Features} & \textit{TS} & \textit{CD} & \textit{} & \textit{Features}  & \textit{TS} & \textit{CD} & \textit{} & \textit{Features}                       & \textit{TS}                    & \textit{CD}          \\ \midrule
SCR duration min* & 4.23        & 0.49        &           & MeanNN*            & 4.59        & 0.39        &           & MeanNN*                                 & 4.99                           & 0.48                 \\
SDNN*             & 3.59        & 0.41        &           & SCR RiseTime mean* & 4.43        & 0.35        &           & WMax*                                   & 4.65                           & 0.44                 \\
C2d*              & 3.43        & 0.39        &           & WMax*              & 3.71        & 3.15        &           & SCR amplitude max*                      & 4.43                           & 0.40                 \\ \bottomrule
% SCR duration mean & 3.3         & 0.38        &           & SCR duration mean* & 3.82        & 0.31        &           & SCR wavelet 4Hz std*                    & 3.9                            & 0.34                 \\
% SD1               & 3.32        & 0.36        &           & MadNN              & 2.87        & 0.27        &           & SCR duration min*                       & 3.4                            & 0.33                 \\ \bottomrule
\end{tabular}%
}
\end{table}

\section{Prediction Modeling}
Primarily, we applied classical machine learning models to the sensor-extracted features (Section \ref{section: feature_extraction}) to classify/predict anxious versus non-anxious participants. These models are well-suited for building prediction models when the dataset size is relatively small, compared to deep learning models. Additionally, this type of modeling is commonly used in mental health research due to its interpretability.

In this work, we have used five well-known classifiers explained as follows. \textbf{Logistic Regression (C1)} \cite{hosmer2013applied}: a statistical model that uses a logistic function to model binary outcomes; 
\textbf{Random Forest (C2)} \cite{breiman2001random}: an ensemble of decision trees that improves classification accuracy through majority voting;
\textbf{Decision Tree (C3)} \cite{myles2004introduction}: a tree-like model used for making decisions based on feature splits;
\textbf{Support Vector Machine (C4)} \cite{burges1998tutorial}: a supervised learning model that finds the hyperplane that best separates classes; and \textbf{Gradient Boosting (C5)} \cite{chen2016xgboost}: that creates a robust model by sequentially adding weak learners (typically decision trees) to minimize prediction errors using gradient descent. Hereafter, we refer to these classifiers as  \textbf{C1 }(Logistic Regression), \textbf{C2} (Random Forest), \textbf{C3} (Decision Tree), \textbf{C4} (Support Vector Machine), and \textbf{C5} (Gradient Boosting).

Furthermore, following recent trends, we also employed a Deep Neural Network (DNN) due to its ability to capture complex patterns. Unlike traditional machine learning models, DNNs allow model weights to be stored and fine-tuned with new data, enabling continuous refinement if the initial performance is suboptimal. We used a network with five dense layers containing 256, 128, 64, 32, and 16 neurons, respectively, followed by an output layer. The model was optimized using the Adam optimizer with a learning rate of 0.0001. Both the network architecture and learning rate were selected empirically.

\subsection{Ground Truth} \label{section: Ground Truth}
% \noindent{\textbf{\textit{\hl{Ground Truth}}}}:
To create the ground truth, we utilized the cumulative sum of scores provided by participants in the PAQ. Each response to the questions was measured on a Likert scale from 1 to 5, where higher values indicated a greater level of anxiety. The cumulative sum of responses across all five questions ranged from 5 to 25, with a score of 5 representing no anxiety and a score of 25 indicating high anxiety. We set a threshold of 15, classifying participants who scored 15 or higher as anxious, while those who scored below 15 were labeled as non-anxious. This threshold was chosen because 15 represents the median score, and if a participant selected a score of 3 for all five questions, it suggested they experienced some anxiety during the activity performance. The same method and threshold were applied consistently across all three activities to maintain homogeneity in our approach. The counts of anxious and non-anxious participants for each activity were (53, 50) in A1, (26, 77) in A2, and (37, 65) in A3, respectively. Table \ref{tab:labels-distribution} provides the count of windows labeled as anxious and non-anxious for each activity.

\subsection{Training and Testing}
% \noindent{\textbf{\textit{Training and Testing}}}:
As discussed in Section \ref{section: feature_extraction}, we computed HRV features from ECG signals, including time domain, frequency domain, non-linear indices, and RSA features for each participant across all activities. Similarly, we extracted EDA features. To evaluate model generalizability, we trained the model on one activity and tested it on the remaining activities, performing all possible train-test combinations (e.g., training on A1 and testing on A2 and A3, training on A2 and testing on A1 and A3, etc).

We performed classification on both individual feature sets and combinations of feature sets. In total, 31 feature combinations were tested. This included five models trained on individual feature sets, ten models using a fusion of two feature sets, ten models with three feature sets, five models with four feature sets, and 1 model using all five feature sets. This approach allowed us to identify the feature combination that produced the best classification metrics.

\subsection{Evaluation Metrics}
% \noindent{\textbf{\textit{Evaluation Metrics}}}:

The performance of classification models is typically evaluated using metrics such as Accuracy, Precision, Recall, F1 score, and AUROC (Area under the Receiver Operating Characteristic Curve). Mental health researchers often use AUROC \cite{fawcett2006introduction} to assess model performance \cite{rashid2020predicting, salekin2018weakly}. However, we found that this approach can be biased when datasets are imbalanced, leading to misleading interpretations. In mental health research, datasets are often imbalanced due to the relatively low number of participants suffering from mental health disorders. For example, we observed that a model trained on a balanced dataset but tested on an imbalanced dataset produced a confusion matrix of [[8, 188], [13, 494]], resulting in an accuracy of 71\% and an AUROC of 0.59, which might seem better than the baseline of 0.5. However, the model had mostly classified data as non-anxious, and due to the larger number of non-anxious instances, the accuracy and AUROC appeared deceptively high. In reality, only 4\% of the anxious class was correctly classified. Even computing average precision and Recall can lead to similarly skewed results. This issue can also occur in reverse, where the model primarily classifies data as anxious, leading to fewer false negatives but an inflated number of false positives. Both scenarios are problematic in mental health research, as misclassifying anxious or non-anxious individuals can have serious consequences.

In this study, instead of relying solely on AUROC or other conventional metrics to evaluate our models, we focused on the Recall of the anxious and non-anxious classes (see definitions in Table \ref{tab:confusion_matrix}). Our main objective was to minimize Type II errors (false negatives) while also keeping Type I errors (false positives) low. We first computed the true positive (anxious correctly classified) and true negative (non-anxious correctly classified) rates by keeping the anxious label as the positive class. Although minimizing Type II errors was our priority, we ensured that Type I errors remained low as well. 

\begin{table*}[]
\centering
\begin{scriptsize}
% Caption spanning across both minipages
\makebox[\textwidth]{\parbox{0.9\textwidth}{\centering
    \caption{Confusion matrix and recall formulas for anxious and non-anxious classification.}
    \label{tab:confusion_matrix}
}}

\vspace{-4pt}
\begin{minipage}{0.5\textwidth}
    \centering
    % \vspace{10pt}
    \begin{tabular}{lcc}
        \toprule
        & \textbf{Predicted anxious} & \textbf{Predicted non-anxious} \\ \midrule
        \textbf{Actual anxious} & True Positive (TP) & False Negative (FN) \\
        \textbf{Actual non-anxious} & False Positive (FP) & True Negative (TN) \\ \bottomrule
    \end{tabular}
\end{minipage}
\hfill
\begin{minipage}{0.45\textwidth}
    % \vspace{2pt} % Fine-tune alignment if needed
    \begin{align*}
        \text{\textbf{Recall}}_{\text{\textit{Anxious}}} &= \frac{\text{True Positives (TP)}}{\text{True Positives (TP)} + \text{False Negatives (FN)}} \\[10pt]
        \text{\textbf{Recall}}_{\text{\textit{Non-anxious}}} &= \frac{\text{True Negatives (TN)}}{\text{True Negatives (TN)} + \text{False Positives (FP)}}
    \end{align*}
\end{minipage}

\end{scriptsize}
\end{table*}

Since we were interested in high Recall of the anxious class but keeping in mind the Recall of the non-anxious class as well, we report the AUROC for models only when at least Recall of the non-anxious label was 50\% while achieving the highest Recall of the anxious class. For instance, if classification model C1 correctly classified 80\% of the anxious label but only 20\% of the non-anxious label, and model C2 correctly classified 65\% of the anxious label and 55\% of the non-anxious label, we would prefer C2. This is because C2 learned the patterns of both the anxious and non-anxious groups, while C1 struggled to recognize the non-anxious cases.

\section{Results (within-participant generalization)}
 In this section, first, we report the above-mentioned classification metrics for within-activity  evaluation. Later, we report the classification metrics for cross-activity evaluation. The classification metrics are represented throughout this paper as a four-item \textit{tuple }consisting of AUROC, Recall (\%) of the anxious class, Recall (\%)  of the non-anxious class, and the best classification model. In both within and cross dataset evaluations, we have used all five classifiers (C1, C2, C3, C4, and C5) for different combinations of features during experimentation. However, for brevity, throughout the paper, we report the results obtained with the best classifiers for all possible feature combinations. 
 
\subsection{Within-activity evaluation} \label{section: within activity within participants}
Within-activity evaluation was performed by following the famous within-dataset evaluation approach.
Within-dataset evaluation refers to using one portion of a dataset for training and the other portion for testing. 
Table \ref{tab:within_dataset} presents the within-activity classification results. We applied a 5-fold stratified cross-validation approach and calculated the mean of the evaluation metrics across all five folds. The stratified 5-fold method was used to maintain class distribution in each training and testing set.

Using ML, we achieved the highest AUROC of 0.82 with the feature set combination \textit{F1 + F5} for Activity 1 using C2 classifier. The recall for the anxious and non-anxious classes was 72.11\% and 73.68\%, respectively. For Activity 2, none of the feature combinations resulted in a recall higher than 50\% for the anxious label. The highest recall achieved was 49.96\% with the combination \textit{F3 + F4 + F5} using C3 classifier, while the highest AUROC was 0.64, and the recall for the anxious class was 46.90\%, and the recall of non-anxious class was 75\%. For Activity 3, the highest AUROC was 0.82 with the feature \textit{F1 + F4 + F5} using C5 classifier. The recall for the anxious class was 56.55\%, while the recall for the non-anxious class was 87.45\%. 

Using DNN, the highest AUROC achieved for Activity 1, Activity 2, and Activity 3 was 0.77, 0.79, and 0.76, respectively, with feature sets \textit{F1 + F2 + F3}, \textit{F1 + F2 + F3 + F5}, and \textit{F1 + F3 + F5} (see Table \ref{app:tab:DNN_within_dataset} in the appendix). Although the AUROC for Activity 2 exceeded the best-performing ML model, the recall for the anxious class was only 0.53. Overall, the DNN underperformed compared to ML across all activities, suggesting that the increased model complexity may not be beneficial for learning patterns from these diverse mental health datasets.

\begin{table*}[]
\centering
\small
\caption{Within-activity (within-dataset) classification evaluation metrics with different combinations of feature sets for Activity 1, Activity 2, and Activity 3. Each tuple represents AUROC, Recall (\%) of the anxious class, Recall (\%)  of the non-anxious class, best classification model. Labels C1, C2, C3, C4, and C5 refer to classifiers Logistic Regression, Random Forest, Decision Tree, Support Vector Machine (SVM), and Gradient Boosting, respectively. Column-wise bold tuples represent the best AUROC performance observed during the within-dataset evaluation. In cases of AUROC ties, the feature set with the highest Anxious Recall was selected.
}
\label{tab:within_dataset}
\begin{tabular}{@{}lccc@{}}
\toprule
\textbf{Feature Set}            & \textbf{Activity 1}       & \textbf{Activity 2}       & \textbf{Activity 3}       \\ \midrule
\textit{Baseline}               & (0.5,0.5,0.5,-)           &(0.5,0.5,0.5,-)            & (0.5,0.5,0.5,-) \\ \midrule
\textit{F1}                     & (0.70, 75.36, 57.93, C4) & (0.61, 41.72, 80.27, C3) & (0.79, 54.79, 85.42, C2) \\
\textit{F2}                     & (0.49, 42.85, 54.56, C5) & (0.51, 23.69, 77.70, C3) & (0.54, 42.97, 65.28, C3) \\
\textit{F3}                     & (0.72, 68.61, 59.46, C2) & (0.56, 34.33, 78.18, C3) & (0.57, 45.51, 68.51, C3) \\
\textit{F4}                     & (0.53, 55.70, 51.08, C3) & (0.53, 31.19, 75.49, C3) & (0.55, 44.44, 65.13, C3) \\
\textit{F5}                     & (0.70, 70.26, 59.34, C2) & (0.54, 33.11, 74.63, C3) & (0.60, 46.21, 73.94, C3) \\ \midrule
\textit{F1 + F2}                & (0.75, 68.52, 69.25, C2) & (0.61, 41.54, 80.98, C3) & (0.64, 58.06, 70.03, C3) \\
\textit{F1 + F3}                & (0.77, 72.14, 64.64, C2) & (0.61, 43.25, 78.86, C3) & (0.61, 48.31, 73.83, C3) \\
\textit{F1 + F4}                & (0.70, 73.02, 58.69, C2) & (0.57, 33.89, 79.20, C3) & (0.71, 50.45, 79.49, C5) \\
\textit{F1 + F5}                & \textbf{(0.82, 72.11, 73.68, C2)} & (0.62, 47.61, 76.17, C3) & (0.64, 54.30, 73.94, C3) \\
\textit{F2 + F3}                & (0.69, 65.41, 63.20, C4) & (0.57, 36.31, 78.43, C3) & (0.58, 48.12, 68.84, C3) \\
\textit{F2 + F4}                & (0.54, 54.94, 50.27, C5) & (0.49, 23.21, 75.46, C3) & (0.54, 46.82, 60.40, C3) \\
\textit{F2 + F5}                & (0.66, 61.43, 60.64, C2) & (0.59, 38.83, 79.60, C3) & (0.58, 45.91, 70.43, C3) \\
\textit{F3 + F4}                & (0.70, 76.30, 52.72, C4) & (0.56, 35.76, 76.24, C3) & (0.56, 51.14, 61.79, C3) \\
\textit{F3 + F5}                & (0.73, 69.30, 65.65, C5) & (0.64, 46.13, 79.90, C3) & (0.61, 48.88, 72.68, C3) \\
\textit{F4 + F5}                & (0.72, 75.78, 54.83, C2) & (0.57, 37.24, 76.18, C3) & (0.61, 51.65, 69.66, C3) \\ \midrule
\textit{F1 + F2 + F3}           & (0.73, 67.65, 66.22, C2) & (0.62, 46.15, 78.63, C3) & (0.60, 50.19, 68.86, C3) \\
\textit{F1 + F2 + F4}           & (0.69, 72.77, 58.45, C2) & (0.57, 34.99, 79.69, C3) & (0.69, 46.05, 78.64, C5) \\
\textit{F1 + F2 + F5}           & (0.79, 71.46, 70.75, C2) & \textbf{(0.64, 46.90, 80.76, C3)} & (0.66, 60.91, 70.18, C3) \\
\textit{F1 + F3 + F4}           & (0.72, 77.33, 55.93, C4) & (0.58, 37.95, 77.36, C3) & (0.63, 57.98, 67.44, C3) \\
\textit{F1 + F3 + F5}           & (0.79, 74.87, 68.28, C2) & (0.64, 46.38, 81.37, C3) & (0.80, 51.14, 90.73, C5) \\
\textit{F1 + F4 + F5}           & (0.80, 78.30, 66.58, C2) & (0.61, 44.90, 77.76, C3) & \textbf{(0.82, 56.55, 87.45, C5)} \\
\textit{F2 + F3 + F4}           & (0.69, 69.37, 57.46, C2) & (0.57, 36.87, 77.32, C3) & (0.56, 47.24, 64.53, C3) \\
\textit{F2 + F3 + F5}           & (0.74, 68.65, 68.61, C2) & (0.63, 47.36, 78.09, C3) & (0.58, 45.00, 71.69, C3) \\
\textit{F2 + F4 + F5}           & (0.70, 75.78, 53.82, C2) & (0.58, 40.27, 76.53, C3) & (0.61, 52.70, 69.96, C3) \\ \midrule
\textit{F3 + F4 + F5}           & (0.75, 75.83, 56.23, C2) & (0.62, 49.96, 75.00, C3) & (0.59, 51.70, 66.66, C3) \\
\textit{F1 + F2 + F3 + F4}      & (0.74, 74.77, 56.83, C2) & (0.59, 38.03, 79.69, C3) & (0.58, 49.55, 66.57, C3) \\
\textit{F1 + F2 + F3 + F5}      & (0.79, 72.59, 68.58, C2) & (0.63, 44.91, 80.74, C3) & (0.63, 55.00, 70.94, C3) \\
\textit{F1 + F2 + F4 + F5}      & (0.78, 78.30, 64.24, C2) & (0.62, 45.41, 79.10, C3) & (0.65, 54.91, 75.91, C3) \\
\textit{F1 + F3 + F4 + F5}      & (0.80, 78.90, 64.53, C2) & (0.63, 48.95, 77.56, C3) & (0.82, 55.44, 87.14, C5) \\
\textit{F2 + F3 + F4 + F5}      & (0.79, 80.15, 63.57, C2) & (0.59, 39.28, 77.92, C3) & (0.60, 50.03, 69.97, C3) \\ \midrule
\textit{F1 + F2 + F3 + F4 + F5} & (0.80, 76.42, 61.45, C2) & (0.62, 41.29, 83.24, C3) & (0.64, 54.88, 73.89, C3) \\ \bottomrule
\end{tabular}
\end{table*}

\subsection{Cross-activity evaluation} \label{section: within participants cross activity}
Cross-activity evaluation was performed by following the famous cross-dataset evaluation approach.
Cross-dataset evaluation refers to using one training dataset and a different testing dataset.  
Following the approach, we trained classifiers on one activity and tested them on others. Table  \ref{tab:in_detail_generalizability_evaluation_metric} shows the classification metrics for the cross-activity evaluation. In the table, the top row represents the training, and the second row shows testing activities. Abbreviations A1, A2, and A3 refer to Activity 1, Activity 2, and Activity 3, respectively. We trained and tested 930 classification models (i.e., five classifiers, 31 feature set combinations, six train-test combinations = 5 x 31 x 6 = 930).

When trained on A1 and tested on A2, we found the highest AUROC of 0.62, with a recall of 59.70\% for the anxious class and 62.21\% for the non-anxious class using the feature combination \textit{F1 + F2 + F3 + F4} with C4 classifier. When tested on A3, the highest AUROC was 0.60 using \textit{F1} with C4. However, when trained on A2 and tested on A1 and A3, none of the feature combinations surpassed the baseline recall of 50\% for the anxious class. While the best AUROC was 0.61 and 0.57 in A1 and A3, respectively, this was mainly due to the large number of non-anxious training samples in A2. Likewise, when trained on A3 and tested on A1 and A2, the highest AUROC scores were 0.65 and 0.59, but these were again dominated by the high recall of the non-anxious class.

To enhance the interpretability of the results in Table \ref{tab:in_detail_generalizability_evaluation_metric}, we created a concise summary in Table \ref{app:tab:compact_evaluation} (see appendix). This compact table shows the mean and standard deviation of AUROC and recall percentages for both anxious and non-anxious classes across six train-test combinations for each feature set (data sourced from Table \ref{tab:in_detail_generalizability_evaluation_metric}). We found that none of the feature combinations exceeded the baseline threshold of 50\% recall for the anxious class, raising questions about cross-dataset generalizability. Specifically, the highest AUROC was 0.59, with a recall of 47.75\% for the anxious class and 67.61\% for the non-anxious class, achieved with the feature combination \textit{F1 + F3 + F4}.

\begin{table*}[]
\centering
\scriptsize
\caption{Cross-activity classification evaluation metrics with different combinations of feature sets. Column labels A1, A2, and A3 correspond to Activity 1, Activity 2, and Activity 3, respectively. Row tuples represent AUROC, Recall (\%) of the anxious class, Recall (\%)  of the non-anxious class, best classification model. Labels C1, C2, C3, C4, and C5 refer to classifiers  Logistic Regression, Random Forest, Decision Tree, Support Vector Machine (SVM), and Gradient Boosting, respectively.
Empty cells with ``-'' means none of the classifiers achieved 50\% recall for the non-anxious class. Column-wise bold tuples represent the best AUROC performance observed during the cross-activity evaluation. In cases of AUROC ties, the feature set with the highest Anxious Recall was selected.}
\label{tab:in_detail_generalizability_evaluation_metric}
\resizebox{\columnwidth}{!}{
\begin{tabular}{@{}llcccccccc@{}}
\toprule
\textbf{$\backslash$ Train}            &  & \multicolumn{2}{c}{\textbf{A1}}               & \multicolumn{1}{c}{} & \multicolumn{2}{c}{\textbf{A2}}               &  & \multicolumn{2}{c}{\textbf{A3}}               \\ \cmidrule{1-1} \cmidrule{3-4} \cmidrule{6-7} \cmidrule{9-10}
\textbf{Feature Set$\backslash$Test} &  & \textit{\textbf{A2}}           & \textit{\textbf{A3}}           &                      & \textit{\textbf{A1}}           & \textit{\textbf{A3}}           &  & \textit{\textbf{A1}}           & \textit{\textbf{A2}}           \\ \midrule
$\backslash$Baseline                  &  & (0.50,50.00,50.00,-)         & (0.50,50.00,50.00,-)         &                      & (0.50,50.00,50.00,-)          & (0.50,50.00,50.00,-)         &  & (0.50,50.00,50.00,-)          &  (0.50,50.00,50.00,-)  \\ \midrule
F1                        &  & (0.59,55.83,57.56,C4) & \textbf{(0.60,55.48,59.84,C4)} &                      & (0.52,33.63,71.24,C3) & (0.50,25.00,74.61,C3) &  & (0.54,46.64,61.80,C3) & (0.55,41.10,68.62,C3) \\
F2                        &  & (0.49,48.31,50.58,C2) & (0.51,50.52,52.27,C2) &                      & (0.50,26.13,73.16,C3) & (0.51,26.80,75.94,C3) &  & (0.51,38.74,62.34,C3) & (0.50,37.85,62.84,C3) \\
F3                        &  & (0.53,51.23,54.25,C3) & (0.55,52.05,57.68,C3) &                      & (0.56,33.18,79.31,C3) & (0.49,20.21,78.74,C3) &  & (0.56,44.84,68.10,C3) & (0.57,44.48,68.73,C3) \\
F4                        &  & (0.52,53.99,50.74,C4) & (0.53,49.60,53.08,C4) &                      & (0.50,29.56,70.11,C3) & (0.50,26.59,72.56,C3) &  & (0.48,34.48,60.87,C3) & (0.53,44.49,62.00,C3) \\
F5                        &  & (0.52,52.26,52.54,C5) & (0.56,56.36,54.70,C3) &                      & (0.49,16.16,81.64,C3) & (0.50,17.80,82.48,C3) &  & (0.50,30.81,70.05,C3) & (0.47,27.44,66.24,C3) \\ \midrule
F1 + F2                   &  & (0.61,55.08,59.27,C4) & (0.60,50.17,63.91,C4) &                      & (0.53,36.04,70.56,C3) & (0.53,28.87,77.12,C3) &  & (0.52,39.19,64.94,C3) & (0.57,33.54,77.03,C2) \\
F1 + F3                   &  & (0.59,51.84,62.64,C4) & (0.59,51.03,61.02,C4) &                      & (0.49,26.91,71.12,C3) & (0.54,30.48,76.97,C3) &  & (0.65,52.47,71.55,C1) & (0.55,47.55,61.58,C3) \\
F1 + F4                   &  & (0.55,54.37,56.44,C3) & (0.56,55.56,57.18,C2) &                      & (0.53,35.96,70.65,C3) & (0.55,30.56,79.49,C3) &  & (0.53,44.83,60.33,C3) & (0.54,45.63,61.88,C3) \\
F1 + F5                   &  & (0.60,58.87,54.82,C1) & (0.57,57.01,56.14,C3) &                      & (0.56,35.75,75.26,C3) & (0.57,34.39,78.70,C3) &  & (0.53,44.13,61.58,C3) & (0.53,38.31,68.42,C3) \\
F2 + F3                   &  & (0.56,44.31,64.89,C2) & (0.52,48.11,56.80,C3) &                      & (0.54,30.63,77.06,C3) & (0.51,23.71,78.30,C3) &  & (0.52,49.55,55.41,C3) & (0.52,44.00,59.67,C3) \\
F2 + F4                   &  & (0.55,54.75,51.67,C2) & (0.53,54.76,50.90,C3) &                      & (0.54,35.15,72.68,C3) & (0.54,28.97,79.43,C3) &  & (0.49,39.11,59.56,C3) & (0.53,48.29,57.13,C3) \\
F2 + F5                   &  & (0.53,55.47,50.22,C3) & (0.51,50.00,52.88,C3) &                      & (0.52,20.67,82.98,C3) & (0.54,30.45,77.94,C3) &  & (0.47,33.52,60.64,C3) & (0.49,36.03,62.56,C3) \\
F3 + F4                   &  & (0.58,56.27,54.95,C4) & (0.56,51.19,57.69,C2) &                      & (0.51,30.54,72.28,C3) & (0.53,25.00,81.79,C3) &  & (0.57,48.77,62.50,C1) & (0.53,46.77,59.78,C3) \\
F3 + F5                   &  & (0.58,62.10,53.37,C3) & (0.57,54.75,54.89,C1) &                      & (0.48,27.37,69.31,C3) & (0.52,29.41,73.68,C3) &  & (0.53,45.81,59.26,C3) & (0.54,41.13,66.42,C3) \\
F4 + F5                   &  & (0.56,61.22,50.39,C3) & (0.52,51.63,52.81,C3) &                      & (0.54,29.81,77.78,C3) & (0.49,22.83,75.58,C3) &  & (0.50,36.02,63.19,C3) & (0.49,39.29,58.86,C3) \\ \midrule
F1 + F2 + F3              &  & (0.60,47.69,66.54,C4) & (0.54,48.11,59.37,C3) &                      & (0.53,30.63,76.19,C3) & (0.52,26.80,77.91,C3) &  & \textbf{(0.65,52.70,71.00,C1)} & (0.55,49.23,60.25,C3) \\
F1 + F2 + F4              &  & (0.57,58.17,52.91,C1) & (0.59,53.97,59.38,C2) &                      & (0.57,37.62,76.50,C3) & (0.50,23.41,76.35,C3) &  & (0.53,45.54,59.56,C3) & (0.55,49.05,61.21,C3) \\
F1 + F2 + F5              &  & (0.58,59.92,56.53,C3) & (0.54,50.00,59.90,C1) &                      & (0.53,34.64,71.28,C3) & (0.55,32.27,77.69,C3) &  & (0.55,49.16,61.17,C3) & (0.53,36.44,68.58,C3) \\
F1 + F3 + F4              &  & (0.61,60.46,57.43,C4) & (0.59,55.16,57.44,C4) &                      & \textbf{(0.61,42.86,78.26,C3)} & (0.56,28.57,82.82,C3) &  & (0.58,54.19,62.50,C1) & (0.59,45.25,67.20,C1) \\
F1 + F3 + F5              &  & (0.51,49.60,52.79,C3) & (0.54,49.32,59.15,C3) &                      & (0.49,26.82,71.43,C3) & (0.51,27.15,74.44,C3) &  & (0.58,56.42,58.73,C3) & (0.56,44.35,68.48,C3) \\
F1 + F4 + F5              &  & (0.59,60.20,56.69,C2) & (0.52,50.54,56.11,C5) &                      & (0.53,39.13,67.36,C3) & \textbf{(0.57,38.59,75.58,C3)} &  & (0.52,50.93,52.78,C3) & (0.52,40.82,63.98,C3) \\
F2 + F3 + F4              &  & (0.60,54.75,60.97,C4) & (0.56,49.60,59.90,C4) &                      & (0.53,32.67,72.68,C3) & (0.54,25.40,82.01,C3) &  & (0.57,49.01,61.75,C1) & (0.55,48.67,61.71,C3) \\
F2 + F3 + F5              &  & (0.56,52.23,56.32,C1) & (0.53,52.73,52.38,C3) &                      & (0.53,30.17,76.06,C3) & (0.53,30.00,75.69,C3) &  & (0.49,40.78,56.91,C3) & (0.51,38.06,64.56,C3) \\
F2 + F4 + F5              &  & -                     & (0.54,55.43,52.48,C3) &                      & (0.50,26.71,73.43,C3) & (0.55,32.61,77.89,C3) &  & (0.52,48.45,55.94,C3) & (0.50,38.27,60.95,C3) \\
F3 + F4 + F5              &  & (0.55,55.10,53.94,C4) & (0.54,51.63,56.11,C1) &                      & (0.50,31.06,69.44,C3) & (0.52,30.98,72.61,C3) &  & (0.53,44.72,57.64,C1) & (0.51,38.27,63.78,C3) \\ \midrule
F1 + F2 + F3 + F4         &  & \textbf{(0.62,59.70,62.21,C4)} & (0.58,50.00,63.24,C4) &                      & (0.57,42.57,71.58,C3) & (0.53,29.76,77.12,C3) &  & (0.59,55.45,61.75,C1) & \textbf{(0.60,44.11,68.03,C1)} \\
F1 + F2 + F3 + F5         &  & (0.57,46.96,64.26,C1) & (0.58,45.91,64.91,C1) &                      & (0.52,30.17,73.40,C3) & (0.52,30.91,73.68,C3) &  & (0.59,51.96,66.49,C3) & (0.52,36.84,67.79,C3) \\
F1 + F2 + F4 + F5         &  & (0.56,60.20,51.87,C3) & (0.53,51.09,57.10,C5) &                      & (0.50,29.81,69.93,C3) & (0.53,28.26,77.23,C3) &  & (0.56,51.55,60.84,C3) & (0.54,41.84,65.88,C3) \\
F1 + F3 + F4 + F5         &  & (0.60,63.78,50.98,C2) & (0.55,52.17,57.43,C3) &                      & (0.54,31.06,77.08,C3) & (0.56,33.15,78.22,C3) &  & (0.57,50.31,63.89,C5) & (0.57,46.94,66.73,C3) \\
F2 + F3 + F4 + F5         &  & (0.56,54.59,54.24,C4) & (0.51,49.46,52.81,C2) &                      & (0.48,27.95,67.83,C3) & (0.53,34.24,72.61,C3) &  & (0.54,47.83,58.04,C1) & (0.51,39.29,62.52,C3) \\
F1 + F2 + F3 + F4 + F5    &  & (0.59,64.29,50.10,C2) & (0.56,48.37,58.75,C2) &                      & (0.56,35.40,76.22,C3) & (0.53,30.43,76.24,C3) &  & (0.60,54.66,66.43,C5) & (0.55,44.39,66.27,C3) \\ \bottomrule
\end{tabular}
}
\end{table*}

\section{Generalizability evaluation on publicly available datasets (RQ2)} \label{cross participants}

In addition to within-activity and cross-activity evaluations on our dataset, we explored other publicly available anxiety datasets collected in controlled settings. Specifically, we included the WESAD dataset \cite{schmidt2018introducing} and the APD dataset \cite{senaratne2021multimodal}, both of which were collected from different geographical regions. This diversity provides a unique opportunity to assess the generalizability of anxiety detection models across varied populations.

\textbf{WESAD (Ws) dataset \cite{schmidt2018introducing}:} The dataset is a wearable, multimodal sensor dataset collected during different emotional states. Data were collected using chest-worn RespiBAN professional sensors, which recorded ECG, EDA, EMG, and temperature at 700 Hz. The study involved 15 participants who underwent different conditions, including baseline, amusement, stress, meditation, and recovery. During the baseline condition, participants read magazines to induce a neutral emotional state for 20 minutes while sensor data were recorded. In the stress condition, participants performed the Trier Social Stress Test (TSST), which involved public speaking and a mental arithmetic task in front of a panel. This task lasted 10 minutes—5 minutes for speaking and 5 minutes for the arithmetic. Following Bhatti et al. \cite{bhatti2024attx}, we labeled the baseline condition as ``non-anxious'' and the TSST condition as ``anxious.'' We used ECG and EDA data from the dataset and applied data cleaning and feature extraction methods as discussed above. The computed features are shown in Table \ref{tab:feature}. Table \ref{tab:labels-distribution} shows the number of data windows labeled as anxious class (AC) and non-anxious class (NAC).

\begin{table*}[]
\centering
\scriptsize
\caption{Number of anxious class (AC) and non-anxious class (NAC) labels in datasets, A1, A2, A3, Ws, and APD for feature sets, F1, F2, F3, F4, and F5. Feature sets F1 to F4 were derived from the ECG signal, while F5 was derived from the EDA signal. There were cases where the ECG signals were faulty while the EDA signals remained intact, or vice versa, leading to an unequal number of samples. Some HRV features also contained NaN or infinite values, which were dropped. As a result, the feature sets have different sample counts.}
\label{tab:labels-distribution}
\begin{tabular}{llcclccccccccccc}
\toprule
\textbf{Dataset} & \textbf{} & \multicolumn{2}{c}{\textbf{A1}} &  & \multicolumn{2}{c}{\textbf{A2}} & \textbf{} & \multicolumn{2}{c}{\textbf{A3}} & \textbf{} & \multicolumn{2}{c}{\textbf{Ws}} & \textbf{} & \multicolumn{2}{c}{\textbf{APD}} \\ 
\cmidrule{3-4} \cmidrule{6-7} \cmidrule{9-10} \cmidrule{12-13} \cmidrule{15-16}
\textbf{Features$\backslash$Count} &  & \textit{\textbf{\#AC}} & \textit{\textbf{\#NAC}} &  & \textit{\textbf{\#AC}} & \textit{\textbf{\#NAC}} & & \textit{\textbf{\#AC}} & \textit{\textbf{\#NAC}} & & \textit{\textbf{\#AC}} & \textit{\textbf{\#NAC}} & & \textit{\textbf{\#AC}} & \textit{\textbf{\#NAC}} \\ 
\midrule
F1 &  & 223 & 233 &  & 326 & 1039 &  & 292 & 508 &  & 394 & 702 &  & 126 & 147 \\
F2 &  & 222 & 231 &  & 325 & 1036 &  & 291 & 507 &  & 394 & 702 &  & 126 & 147 \\
F3 &  & 223 & 232 &  & 326 & 1036 &  & 292 & 508 &  & 394 & 694 &  & 126 & 147 \\
F4 &  & 203 & 184 &  & 263 & 808 &  & 252 & 390 &  & 341 & 344 &  & 114 & 129 \\
F5 &  & 198 & 207 &  & 266 & 788 &  & 236 & 468 &  & 394 & 702 &  & 122 & 151 \\ \midrule
Mean ($\sigma$) &  & 214 (12)& 217 (19)&  & 301 (34)& 941 (117)&  & 273 (27)& 476 (46)&  & 383 (24)& 629 (142)&  & 123 (5)& 144 (8)\\
% STD &  & 12 & 19 &  & 34 & 117 &  & 27 & 46 &  & 24 & 142 &  & 5 & 8 \\ 
\bottomrule
\end{tabular}
\end{table*}

\begin{figure*}[!t]
    \centering
    % First Row: F1, F2, F3
    \begin{minipage}[b]{0.32\textwidth}
        \centering
        \subcaptionbox{F1 \label{fig:TSNE_F1}}{
            \includegraphics[width=\linewidth]{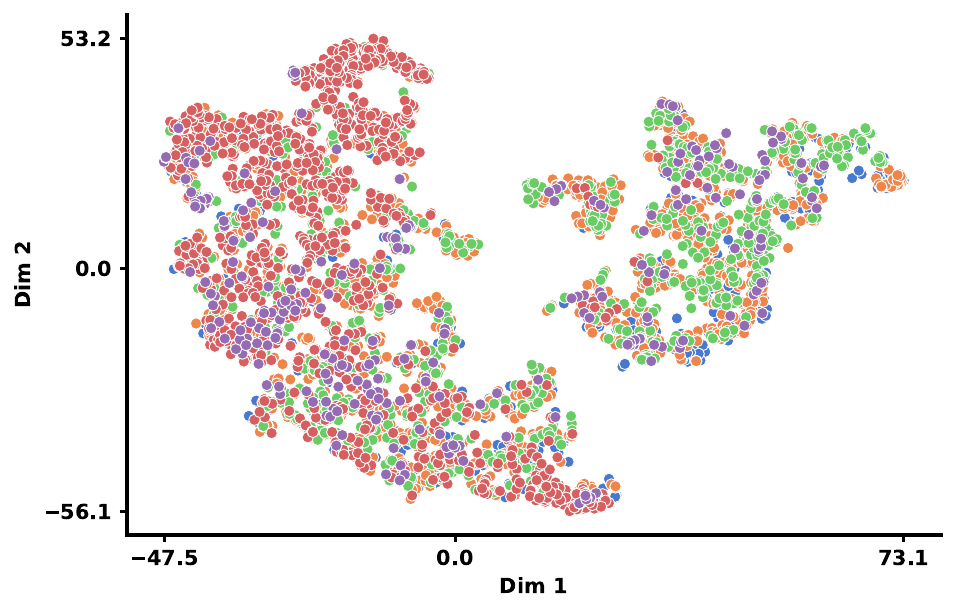}
        }
    \end{minipage}
    \begin{minipage}[b]{0.32\textwidth}
        \centering
        \subcaptionbox{F2 \label{fig:TSNE_F2}}{
            \includegraphics[width=\linewidth]{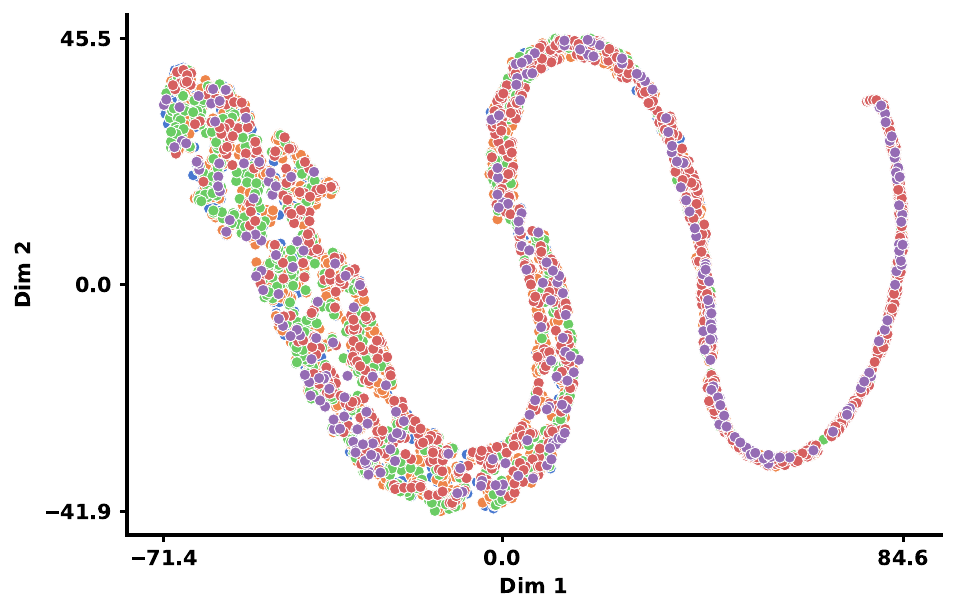}
        }
    \end{minipage}
    \begin{minipage}[b]{0.32\textwidth}
        \centering
        \subcaptionbox{ F3 \label{fig:TSNE_F3}}{
            \includegraphics[width=\linewidth]{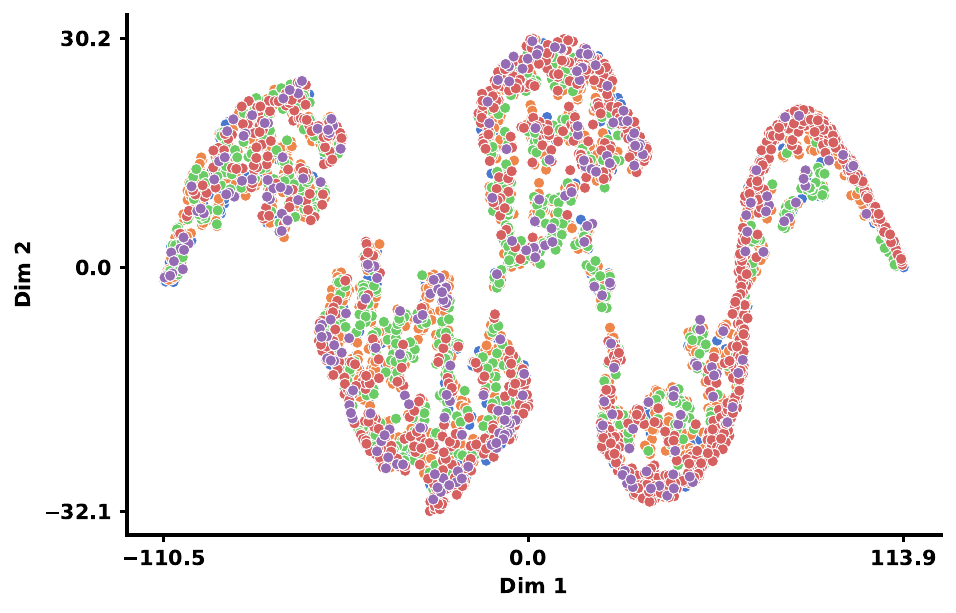}
        }
    \end{minipage}
    
    \vspace{0.5cm} % Space between rows

    % Second Row: F4, F5, Legend
    \begin{minipage}[b]{0.32\textwidth}
        \centering
        \subcaptionbox{F4 \label{fig:TSNE_F4}}{
            \includegraphics[width=\linewidth]{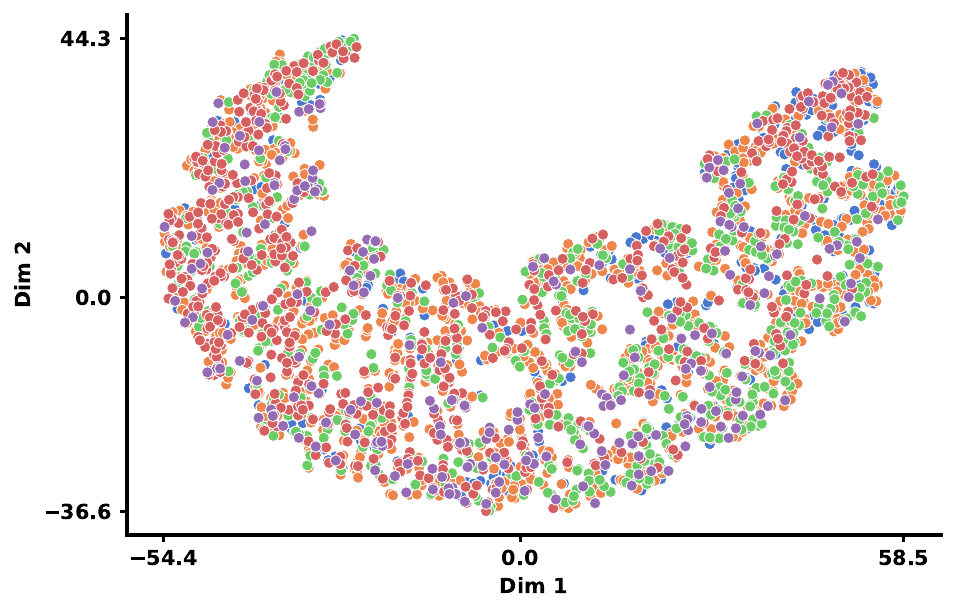} 
        }
    \end{minipage}
    \begin{minipage}[b]{0.32\textwidth}
        \centering
        \subcaptionbox{F5 \label{fig:TSNE_F5}}{
            \includegraphics[width=\linewidth]{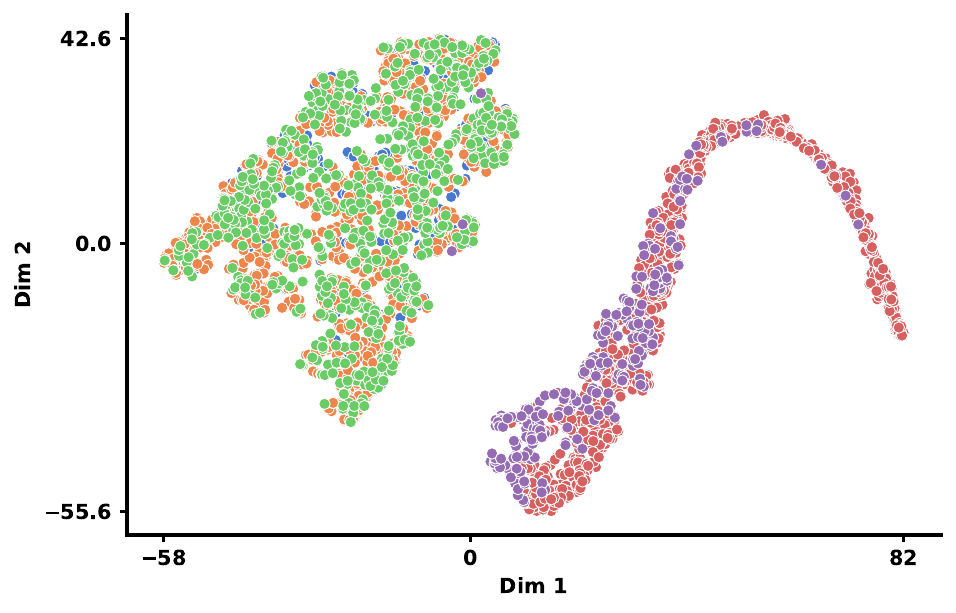} 
        }
    \end{minipage}
    \begin{minipage}[b]{0.32\textwidth}
        \centering
            \includegraphics[width=\linewidth]{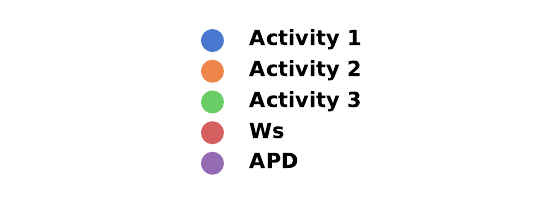} 
            \phantomsubcaption
    \end{minipage}

    \caption{t-SNE plots, i.e., low-dimensional representation of high-dimensional features sets \textbf{(a)} F1, \textbf{(b)} F2, \textbf{(c)} F3, \textbf{(d)} F4, and  \textbf{(e)} F5 collected during Activity 1 (blue), Activity 2 (orange), Activity 3 (green), Ws (maroon), and APD (purple). \textbf{Best viewed in colour.}}
    \Description{}
    \label{fig: TSNE_public}
\end{figure*}

\begin{figure}
    \centering
    \includegraphics[width=0.5\linewidth]{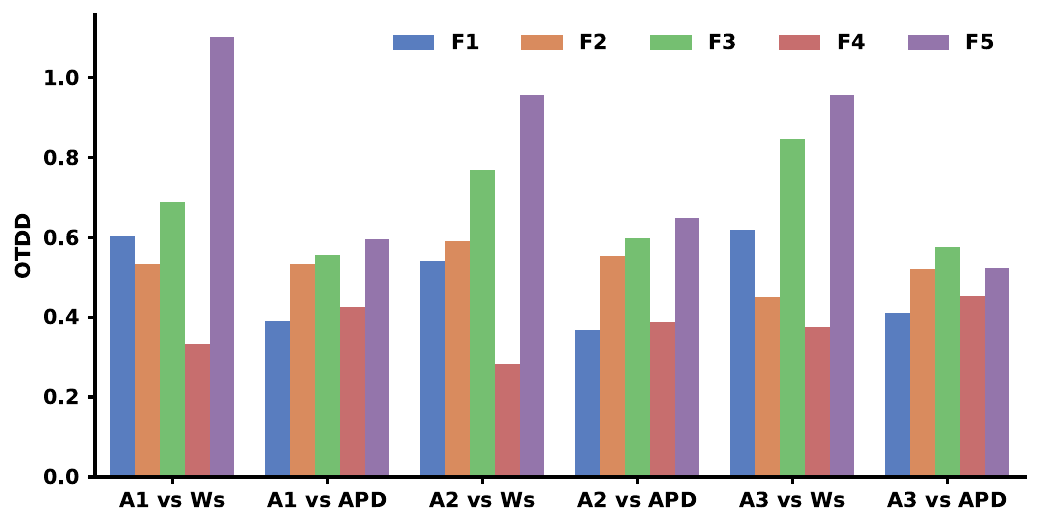}
    \caption{Optimal Transport Dataset Distance (OTDD) values computed between different pairs of datasets (A1, A2, A3, Ws, and APD) on five different feature sets (F1, F2, F3, F4, and F5). \textbf{Best viewed in colour.}}
    \Description{}
    \label{fig:otdd_public}
\end{figure}

\textbf{Anxiety Phase Dataset (APD) \cite{senaratne2021multimodal}:} The dataset contains rich data from multiple sensors, including ECG, EDA, movement, posture, and speech, collected from 95 young adults during two anxiety-inducing experiments. Initially, participants rested for two minutes during a baseline period. They then underwent two anxiety-provoking tasks: one involving exposure to bugs and another requiring public speaking for 3 minutes. ECG was collected using the Zephyr BioHarness 3.0 biopatch, worn on the chest, at 250 Hz. EDA  was recorded with the Grove-GSR Sensor V1.2, worn on the wrist with electrodes on the index and ring fingers, at 50 Hz. After the anxiety-inducing activities, participants used Subjective Units of Distress Scale (SUDS) to rate their distress on a scale from 0 (totally relaxed) to 100 (the highest anxiety or discomfort they have ever felt). We focused on the public speaking task for generalizability, as it aligns with one of the tasks in our study. We used the participants' SUDS scores to label them as ``anxious'' or ``non-anxious.'' Afterward, we applied the same data cleaning and feature extraction methods as mentioned above, with the features listed in Table \ref{tab:feature}. Table \ref{tab:labels-distribution} shows the number of data windows labeled as AC and NAC in our, Ws, and APD dataset.

\subsection{Similarity between datasets} 
Before directly training and testing the classifiers using the publicly available datasets, we followed the technique discussed in Section \ref{section: data_similarity} to check the similarity between different datasets. We plotted t-SNE and computed the OTDD between the Ws and our A1, A2, and A3 datasets, as well as APD and our A1, A2, and A3 datasets. The t-SNE plot in Figure \ref{fig: TSNE_public} shows that  Ws and APD datasets share characteristics similar to ours (A1, A2, and A3) for features \textit{F1, F2, F3}, and \textit{F4}. However, for \textit{F5}, which represents the EDA features, both Ws and APD show different characteristics compared to our data.

To confirm the t-SNE observations, we computed OTDD between different feature sets of all dataset pairs. Figure \ref{fig:otdd_public} presents the OTDD values computed between Ws and our dataset, as well as between APD and our dataset. We found the highest distance between Ws and our dataset for features \textit{F5} and \textit{F3}. Similarly, the APD dataset showed the same pattern, though the distances were smaller than those between Ws and our dataset.

%-----------------------------------------------------------------------------------
\begin{table*}[]
\centering 
\scriptsize
\caption{
Cross-dataset classification evaluation metrics on training on Activity 1 (A1), Activity 2 (A2), and Activity 3 (A3), and testing on WESAD (Ws) and APD with different combinations of feature sets.  Row tuples represent AUROC, Recall (\%) of the anxious class, Recall (\%)  of the non-anxious class, best classification model. Labels C1, C2, C3, C4, and C5 refer to classifiers Logistic Regression, Random Forest, Decision Tree, SVM, and Gradient Boosting, respectively.
Empty cells with ``-'' means none of the classifiers achieved 50\% recall for the non-anxious class. Column-wise bold tuples represent the best AUROC performance observed under different cross-dataset combinations.  In cases of AUROC ties, the feature set with the highest Anxious Recall was selected.}
\label{tab:cross_gen_train_own_test_public}
\resizebox{\columnwidth}{!}{
\begin{tabular}{@{}llcclcclcc@{}}
\toprule
\textbf{$\backslash$Train}            &  & \multicolumn{2}{c}{\textbf{A1}}               & \multicolumn{1}{c}{} & \multicolumn{2}{c}{\textbf{A2}}               &  & \multicolumn{2}{c}{\textbf{A3}}               \\ \cmidrule{1-1} \cmidrule{3-4} \cmidrule{6-7} \cmidrule{9-10}
\textbf{Feature Set$\backslash$Test}       & \textbf{} & \textit{\textbf{Ws} }             & \textit{\textbf{APD} }             &  & \textit{\textbf{Ws}}              & \textit{\textbf{APD}}              &  & \textit{\textbf{Ws}}              & \textit{\textbf{APD}}             \\ \midrule
Baseline                  &  & (0.5,50,50,-)         & (0.5,50,50,-)         &                      & (0.5,50,50,-)         & (0.5,50,50,-)         &  & (0.5,50,50,-)         &  (0.5,50,50,-) \\ \midrule
\textit{F1}                     &           & \textbf{(0.74, 58.88, 77.92, C2)} & \textbf{(0.62, 45.24, 67.35, C1)} &  & (0.44, 26.14, 62.39, C3) & (0.46, 18.25, 73.47, C3) &  & (0.58, 30.2, 85.47, C3)  & (0.52, 30.16, 73.47, C3) \\
\textit{F2}                     &           & (0.54, 55.33, 51.57, C1) & (0.52, 52.38, 56.46, C2) &  & (0.56, 30.46, 81.05, C3) & (0.52, 28.57, 75.51, C3) &  & (0.49, 33.25, 65.38, C3) & (0.49, 34.13, 63.27, C3) \\
\textit{F3}                     &           & (0.58, 54.06, 61.38, C3) & (0.49, 33.33, 63.95, C3) &  & (0.45, 29.7, 60.52, C3)  & (0.5, 26.98, 72.79, C3)  &  & (0.51, 31.47, 71.18, C3) & (0.54, 35.71, 72.79, C3) \\
\textit{F4}                     &           & (0.7, 78.01, 50.29, C1)  & (0.55, 58.77, 51.94, C3) &  & (0.52, 27.86, 75.29, C3) & (0.54, 28.95, 79.07, C3) &  & (0.5, 34.6, 65.12, C3)   & (0.47, 34.21, 60.47, C3) \\
\textit{F5}                     &           & (0.5, 0.0, 100.0, C4)    & (0.54, 51.64, 56.29, C1) &  & (0.47, 40.61, 59.83, C1) & (0.57, 45.9, 66.89, C1)  &  & (0.5, 38.58, 62.39, C3)  & (0.56, 60.66, 57.62, C5) \\ \midrule
\textit{F1 + F2}                &           & (0.7, 64.72, 61.11, C2)  & (0.6, 50.0, 62.59, C1)   &  & (0.49, 29.95, 68.66, C3) & (0.47, 23.81, 70.07, C3) &  & (0.51, 29.44, 72.65, C3) & (0.53, 35.71, 70.07, C3) \\
\textit{F1 + F3}                &           & (0.7, 54.06, 77.52, C2)  & (0.52, 38.1, 65.31, C3)  &  & (0.49, 30.2, 66.86, C3)  & (0.48, 32.54, 63.27, C3) &  & (0.53, 45.69, 61.1, C3)  & (0.54, 44.44, 63.27, C3) \\
\textit{F1 + F4}                &           & (0.65, 64.81, 57.56, C5) & (0.51, 42.11, 71.32, C5) &  & (0.52, 29.03, 74.13, C3) & (0.53, 21.93, 84.5, C3)  &  & (0.51, 36.66, 64.53, C3) & (0.54, 37.72, 69.77, C3) \\
\textit{F1 + F5}                &           & (0.5, 0.0, 100.0, C4)    & (0.53, 50.42, 56.46, C1) &  & (0.73, 63.71, 74.07, C2) & (0.53, 52.1, 53.06, C3)  &  & (0.5, 0.0, 100.0, C4)    & (0.6, 42.86, 76.19, C1)  \\
\textit{F2 + F3}                &           & (0.68, 55.58, 68.73, C2) & (0.48, 41.27, 54.42, C3) &  & (0.47, 28.17, 65.27, C3) & (0.47, 24.6, 70.07, C3)  &  & (0.48, 31.47, 63.54, C3) & (0.55, 41.27, 68.03, C3) \\
\textit{F2 + F4}                &           & (0.6, 48.09, 65.7, C4)   & (0.55, 54.39, 52.71, C2) &  & (0.49, 23.17, 75.0, C3)  & (0.55, 27.19, 82.17, C3) &  & (0.7, 31.38, 91.57, C1)  & (0.48, 35.96, 60.47, C3) \\
\textit{F2 + F5}                &           & (0.5, 0.0, 100.0, C4)    & (0.56, 52.94, 58.5, C1)  &  & (0.48, 38.83, 58.4, C1)  & (0.61, 50.42, 65.99, C5) &  & (0.5, 34.26, 66.67, C3)  & (0.6, 47.9, 72.11, C1)   \\
\textit{F3 + F4}                &           & (0.6, 56.3, 61.05, C2)   & (0.49, 47.37, 55.81, C4) &  & (0.51, 32.26, 70.35, C3) & (0.53, 31.58, 75.19, C3) &  & (0.49, 35.48, 62.21, C3) & (0.55, 46.49, 62.79, C3) \\
\textit{F3 + F5}                &           & (0.63, 55.58, 62.25, C2) & (0.55, 52.1, 58.5, C1)   &  & (0.45, 40.36, 59.65, C1) & (0.55, 54.62, 59.18, C2) &  & (0.5, 0.0, 100.0, C4)    & (0.55, 45.38, 64.63, C3) \\
\textit{F4 + F5}                &           & (0.5, 0.0, 100.0, C4)    & (0.57, 53.27, 60.47, C1) &  & (0.62, 65.98, 53.2, C2)  & (0.57, 50.47, 58.14, C2) &  & -                        & (0.58, 41.12, 74.42, C1) \\ \midrule
\textit{F1 + F2 + F3}           &           & (0.64, 58.38, 69.45, C3) & (0.54, 44.44, 63.95, C3) &  & (0.48, 32.74, 64.12, C3) & (0.51, 23.02, 78.23, C3) &  & (0.56, 46.7, 65.13, C3)  & (0.52, 43.65, 61.22, C3) \\
\textit{F1 + F2 + F4}           &           & (0.54, 49.27, 59.59, C3) & (0.52, 51.75, 50.39, C1) &  & (0.45, 32.26, 57.27, C3) & (0.5, 26.32, 74.42, C3)  &  & \textbf{(0.73, 35.19, 91.57, C1)} & (0.46, 25.44, 66.67, C3) \\
\textit{F1 + F2 + F5}           &           & (0.5, 0.0, 100.0, C4)    & (0.54, 51.26, 56.46, C1) &  & (0.47, 41.62, 59.69, C1) & (0.52, 47.9, 55.1, C3)   &  & (0.5, 0.0, 100.0, C4)    & (0.59, 46.22, 72.11, C1) \\
\textit{F1 + F3 + F4}           &           & (0.67, 56.6, 74.71, C2)  & (0.44, 37.72, 51.16, C3) &  & (0.45, 31.38, 59.59, C3) & (0.48, 44.74, 51.16, C3) &  & (0.46, 35.78, 55.52, C3) & (0.53, 44.74, 62.02, C3) \\
\textit{F1 + F3 + F5}           &           & (0.64, 63.71, 54.47, C2) & (0.6, 56.3, 57.14, C5)   &  & (0.58, 38.83, 78.1, C5)  & (0.63, 49.58, 74.15, C5) &  & (0.5, 0.0, 100.0, C4)    & (0.59, 42.02, 75.51, C1) \\
\textit{F1 + F4 + F5}           &           & (0.65, 59.24, 61.05, C2) & (0.54, 52.34, 56.59, C1) &  & (0.73, 67.45, 78.78, C3) & (0.6, 69.16, 51.16, C3)  &  & (0.5, 0.0, 100.0, C4)    & (0.59, 43.93, 73.64, C1) \\
\textit{F2 + F3 + F4}           &           & (0.64, 44.28, 71.51, C4) & (0.51, 42.11, 67.44, C4) &  & (0.5, 28.74, 71.51, C3)  & (0.5, 34.21, 66.67, C3)  &  & (0.46, 38.12, 54.07, C3) & (0.57, 50.0, 64.34, C3)  \\
\textit{F2 + F3 + F5}           &           & (0.5, 0.0, 100.0, C4)    & (0.56, 52.1, 59.86, C1)  &  & (0.45, 40.61, 56.48, C1) & (0.6, 47.06, 72.79, C1)  &  & (0.51, 30.46, 70.89, C3) & \textbf{(0.65, 57.14, 74.15, C5)} \\
\textit{F2 + F4 + F5}           &           & (0.58, 48.97, 65.7, C2)  & (0.52, 54.21, 51.16, C2) &  & (0.54, 52.2, 50.58, C1)  & (0.62, 53.27, 60.47, C2) &  & -                        & (0.57, 62.62, 51.94, C3) \\
\textit{F3 + F4 + F5}           &           & (0.71, 49.85, 81.4, C2)  & (0.56, 51.4, 60.47, C1)  &  & (0.55, 51.91, 58.43, C3) & (0.57, 53.27, 61.24, C1) &  & (0.5, 0.0, 100.0, C4)    & (0.59, 46.73, 72.09, C1) \\ \midrule
\textit{F1 + F2 + F3 + F4}      &           & (0.69, 44.57, 78.78, C4) & (0.54, 38.6, 71.32, C4)  &  & (0.41, 23.75, 57.56, C3) & (0.54, 23.68, 83.72, C3) &  & (0.47, 31.96, 62.79, C3) & (0.53, 40.35, 65.12, C3) \\
\textit{F1 + F2 + F3 + F5}      &           & (0.67, 40.1, 85.59, C2)  & (0.56, 53.78, 59.18, C1) &  & (0.62, 40.86, 80.98, C5) & (0.59, 45.38, 68.71, C5) &  & (0.5, 0.0, 100.0, C4)    & (0.6, 47.06, 72.79, C1)  \\
\textit{F1 + F2 + F4 + F5}      &           & (0.67, 56.89, 75.58, C2) & (0.55, 52.34, 58.14, C1) &  & \textbf{(0.74, 69.21, 78.49, C3)} & (0.64, 61.68, 55.04, C5) &  & (0.5, 0.0, 100.0, C4)    & (0.59, 45.79, 71.32, C1) \\
\textit{F1 + F3 + F4 + F5}      &           & (0.72, 55.13, 79.65, C2) & (0.57, 53.27, 60.47, C1) &  & (0.61, 43.7, 71.51, C2)  & (0.56, 52.34, 58.91, C1) &  & (0.5, 0.0, 100.0, C4)    & (0.59, 46.73, 71.32, C1) \\
\textit{F2 + F3 + F4 + F5}      &           & (0.73, 58.65, 75.0, C2)  & (0.58, 54.21, 62.02, C1) &  & (0.54, 47.8, 59.01, C5)  & \textbf{(0.67, 68.22, 55.81, C5)} &  & (0.5, 0.0, 100.0, C4)    & (0.59, 46.73, 70.54, C1) \\ \midrule
\textit{F1 + F2 + F3 + F4 + F5} &           & (0.65, 68.04, 51.74, C2) & (0.57, 52.34, 61.24, C1) &  & (0.53, 48.39, 56.98, C5) & (0.57, 52.34, 61.24, C1) &  & (0.5, 0.0, 100.0, C4)    & (0.59, 45.79, 71.32, C1) \\ \bottomrule
\end{tabular}
}
\end{table*}

%----------------------------------------------------------------------
\begin{table*}[]
\centering
\scriptsize
\caption{Cross-dataset classification evaluation metrics on training on  WESAD (Ws) and Multimodel (APD) and testing on 
 Activity 1 (A1), Activity 2 (A2), and Activity 3 (A3) with different combinations of feature sets.  Row tuples represent AUROC, Recall (\%) of the anxious class, Recall (\%)  of the non-anxious class, best classification model. Labels C1, C2, C3, C4, and C5 refer to classifiers Logistic Regression, Random Forest, Decision Tree, SVM, and Gradient Boosting, respectively.
Empty cells with ``-'' means none of the classifiers achieved 50\% recall for the non-anxious class. Column-wise bold tuples represent the best AUROC performance observed under different cross-dataset combinations. In cases of AUROC ties, the feature set with the highest Anxious Recall was selected.}
\label{tab:cross_gen_train_public_test_own}
\begin{tabular}{@{}llccclccc@{}}
\toprule
\textbf{$\backslash$Train}                  & \textbf{} & \multicolumn{3}{c}{\textbf{Ws}}                                                & \multicolumn{1}{c}{\textbf{}} & \multicolumn{3}{c}{\textbf{APD}}                                                \\ \cmidrule{3-5} \cmidrule{7-9}
\textbf{Feature Set$\backslash$Test}       & \textbf{} & \textit{\textbf{A1}}              & \textit{\textbf{A2}}              & \textit{\textbf{A3}}              & \textbf{}                     & \textit{\textbf{A1}}              & \textit{\textbf{A2} }             & \textit{\textbf{A3}}              \\ \midrule

Baseline                        &           & (0.5, 50.0, 50.0, -)        & (0.5, 50.0, 50.0, -)        & (0.5, 50.0, 50.0, -)        &                               & (0.5, 50.0, 50.0, -)        & (0.5, 50.0, 50.0, -)        & (0.5, 50.0, 50.0, -) \\ \midrule
\textit{F1}                     &           & -                        & -                        & -                        &                               & (0.55, 47.98, 60.09, C4) & (0.57, 58.59, 51.97, C1) & (0.59, 60.27, 53.35, C1) \\
\textit{F2}                     &           & (0.49, 50.0, 54.98, C5)  & (0.54, 50.77, 53.76, C2) & (0.52, 49.83, 51.28, C2) &                               & (0.52, 53.6, 51.08, C3)  & (0.54, 54.46, 53.47, C3) & (0.51, 43.99, 57.99, C3) \\
\textit{F3}                     &           & -                        & -                        & -                        &                               & (0.54, 44.84, 61.21, C1) & (0.54, 47.55, 60.42, C5) & (0.52, 48.97, 55.71, C3) \\
\textit{F4}                     &           & -                        & -                        & (0.56, 56.75, 50.26, C2) &                               & -                        & (0.61, 61.22, 62.5, C1)  & (0.54, 57.54, 50.51, C3) \\
\textit{F5}                     &           & (0.57, 18.69, 86.96, C4) & (0.55, 29.32, 79.44, C1) & (0.42, 14.41, 74.79, C4) &                               & (0.62, 31.82, 79.71, C4) & (0.58, 46.62, 69.04, C1) & (0.42, 16.95, 72.22, C4) \\ \midrule
\textit{F1 + F2}                &           & -                        & -                        & -                        &                               & -                        & (0.6, 62.15, 55.21, C1)  & (0.58, 56.7, 55.23, C1)  \\
\textit{F1 + F3}                &           & -                        & -                        & -                        &                               & (0.55, 54.26, 53.88, C1) & (0.55, 57.36, 53.09, C3) & (0.57, 55.14, 58.07, C3) \\
\textit{F1 + F4}                &           & -                        & -                        & -                        &                               & -                        & (0.56, 48.29, 58.04, C4) & (0.52, 52.38, 54.1, C4)  \\
\textit{F1 + F5}                &           & (0.48, 14.53, 80.53, C3) & (0.57, 50.81, 58.77, C2) & (0.59, 34.39, 75.69, C2) &                               & \textbf{(0.64, 74.3, 53.16, C1) } & (0.62, 60.89, 52.34, C1) & (0.58, 56.11, 60.15, C1) \\
\textit{F2 + F3}                &           & -                        & -                        & -                        &                               & (0.52, 38.74, 67.53, C1) & (0.56, 47.08, 61.32, C2) & (0.51, 47.42, 54.64, C3) \\
\textit{F2 + F4}                &           & -                        & -                        & -                        &                               & -                        & (0.58, 65.02, 51.55, C3) & (0.55, 51.98, 54.5, C1)  \\
\textit{F2 + F5}                &           & (0.53, 48.6, 53.72, C4)  & (0.55, 59.11, 51.4, C4)  & (0.5, 33.64, 65.41, C1)  &                               & (0.56, 49.72, 60.64, C1) & (0.63, 66.8, 53.6, C1)   & (0.46, 38.64, 56.64, C1) \\
\textit{F3 + F4}                &           & -                        & (0.5, 47.15, 52.6, C3)   & -                        &                               & (0.53, 49.75, 56.52, C1) & (0.55, 58.17, 51.49, C3) & (0.55, 48.02, 58.46, C2) \\
\textit{F3 + F5}                &           & \textbf{(0.63, 65.92, 56.61, C1)} & (0.55, 53.23, 55.28, C1) & (0.54, 50.23, 55.14, C1) &                               & (0.53, 50.84, 55.03, C3) & (0.57, 57.26, 56.45, C4) & (0.52, 39.82, 57.64, C4) \\
\textit{F4 + F5}                &           & (0.55, 26.09, 78.47, C2) & \textbf{(0.62, 64.29, 55.12, C1) }& (0.56, 53.8, 58.75, C1)  &                               & (0.56, 58.39, 52.08, C1) & (0.64, 57.14, 62.2, C1)  & (0.53, 38.59, 64.03, C1) \\ \midrule
\textit{F1 + F2 + F3}           &           & -                        & -                        & -                        &                               & (0.53, 39.19, 63.64, C1) & (0.58, 54.77, 55.51, C2) & (0.57, 45.7, 64.5, C2)   \\
\textit{F1 + F2 + F4}           &           & -                        & -                        & -                        &                               & -                        & (0.62, 67.3, 50.06, C1)  & (0.54, 56.75, 51.41, C2) \\
\textit{F1 + F2 + F5}           &           & (0.59, 12.29, 92.55, C5) & (0.57, 63.56, 50.22, C2) & (0.58, 50.45, 64.16, C2) &                               & (0.6, 67.04, 53.72, C4)  & (0.64, 63.56, 51.84, C1) & (0.58, 55.0, 56.89, C1)  \\
\textit{F1 + F3 + F4}           &           & -                        & -                        & -                        &                               & (0.53, 31.03, 69.02, C1) & (0.6, 58.56, 56.93, C2)  & (0.57, 50.79, 58.97, C5) \\
\textit{F1 + F3 + F5}           &           & -                        & (0.58, 58.87, 56.01, C2) & (0.59, 57.47, 58.15, C2) &                               & (0.56, 50.84, 57.67, C1) & (0.58, 55.65, 58.8, C2)  & (0.59, 43.44, 71.43, C2) \\
\textit{F1 + F4 + F5}           &           & (0.51, 3.11, 96.53, C5)  & (0.6, 60.2, 52.95, C2)   & (0.59, 65.22, 51.16, C2) &                               & -                        & (0.57, 52.04, 55.51, C4) & (0.56, 53.8, 54.79, C4)  \\
\textit{F2 + F3 + F4}           &           & -                        & -                        & -                        &                               & (0.53, 39.11, 67.21, C1) & (0.6, 58.56, 56.51, C2)  & (0.54, 46.43, 60.41, C2) \\
\textit{F2 + F3 + F5}           &           & (0.61, 58.66, 58.51, C1) & (0.54, 47.77, 57.65, C1) & (0.54, 44.09, 59.65, C1) &                               & (0.52, 51.96, 52.66, C2) & (0.59, 59.92, 56.18, C4) & (0.52, 45.45, 55.89, C4) \\
\textit{F2 + F4 + F5}           &           & (0.56, 33.54, 73.43, C2) & (0.58, 31.63, 77.91, C2) & (0.6, 40.22, 73.6, C2)   &                               & (0.56, 52.17, 54.55, C2) & \textbf{(0.67, 73.47, 50.69, C1)} & (0.57, 50.0, 58.42, C5)  \\
\textit{F3 + F4 + F5}           &           & (0.52, 32.3, 63.19, C2)  & (0.57, 38.78, 69.88, C2) & (0.48, 39.67, 56.11, C3) &                               & (0.52, 41.61, 61.81, C1) & (0.6, 59.69, 53.94, C2)  & (0.59, 49.46, 62.38, C5) \\ \midrule
\textit{F1 + F2 + F3 + F4}      &           & -                        & -                        & -                        &                               & (0.53, 25.25, 73.22, C1) & (0.6, 61.6, 53.53, C2)   & (0.58, 55.16, 60.67, C3) \\
\textit{F1 + F2 + F3 + F5}      &           & -                        & (0.58, 57.49, 52.79, C2) & (0.6, 60.45, 55.14, C5)  &                               & (0.53, 39.11, 63.3, C1)  & (0.6, 54.66, 59.71, C2)  & (0.58, 45.45, 65.16, C2) \\
\textit{F1 + F2 + F4 + F5}      &           & (0.48, 1.24, 97.9, C5)   & (0.59, 4.08, 97.44, C5)  & \textbf{(0.62, 63.04, 55.45, C2)} &                               & (0.53, 58.39, 50.35, C4) & (0.61, 61.73, 50.69, C5) & \textbf{(0.63, 73.91, 51.16, C3)} \\
\textit{F1 + F3 + F4 + F5}      &           & (0.47, 9.32, 84.72, C3)  & (0.57, 10.71, 91.14, C5) & (0.51, 11.41, 86.47, C5) &                               & (0.52, 34.16, 69.44, C1) & (0.61, 66.33, 52.36, C2) & (0.61, 55.98, 58.75, C2) \\
\textit{F2 + F3 + F4 + F5}      &           & (0.53, 53.42, 57.34, C2) & (0.57, 41.33, 69.23, C2) & (0.58, 60.87, 52.81, C2) &                               & (0.51, 36.02, 67.13, C1) & (0.63, 62.24, 59.17, C4) & (0.6, 57.61, 57.1, C5)   \\ \midrule
\textit{F1 + F2 + F3 + F4 + F5} &           & (0.48, 10.56, 84.62, C3) & (0.53, 3.06, 91.91, C5)  & (0.49, 7.07, 87.46, C5)  &                               & (0.51, 27.95, 72.73, C1) & (0.61, 59.18, 52.07, C2) & (0.58, 59.24, 56.77, C3) \\ \bottomrule
\end{tabular}%
\end{table*}
\subsection{Results (Cross-participant generalization)}

In this section, we first report classification results on training on A1, A2, and A3 and testing on Ws and APD. Next, we report the classification results on training Ws and APD and testing on A1, A2, and A3.
 
\subsubsection{ Training on A1, A2, and A3 and testing on Ws and APD}
Table \ref{tab:cross_gen_train_own_test_public} shows the best results obtained across different feature combinations when classifiers were trained on A1, A2, and A3 and tested on the Ws and APD  datasets. When the model was trained on A1, the best results were achieved using F1 for both Ws and APD, with AUROCs of 0.74 and 0.62, respectively. The recall for the anxious and non-anxious classes was 58.88\% and 77.92\% for Ws, and 45.24\% and 67.35\% for APD. When trained on A2, the best AUROC for Ws was 0.74, with a recall of 69.21\% for the anxious class and 78.49\% for the non-anxious class. For APD, the best AUROC was 0.67, with recall of 68.22\% and 55.81\% for the anxious and non-anxious classes, respectively. When trained on A3, none of the results surpassed the baseline recall for the anxious class on Ws. Although the AUROC reached 0.7, it was mainly due to the 91.57\% recall for the non-anxious class. When tested on APD, the best AUROC was 0.65, with a recall of 57.14\% for the anxious class and 74.15\% for the non-anxious class.

Table \ref{app:tab:compact_evaluation_1} in the appendix presents the mean and standard deviation for AUROC and recall percentages for anxious and non-anxious classes across the six train-test combinations (e.g., train A1, test Ws, and APD) for each feature set (data taken from Table \ref{tab:cross_gen_train_own_test_public}). We found that none of the feature combinations exceeded the baseline threshold of 50\% recall for the anxious class, raising questions about the cross-data generalizability with another group of participants. Specifically, the highest AUROC was 0.62, with a recall of 47.65\% for the anxious class and 73.10\% for the non-anxious class, achieved with the feature combination \textit{F1 + F2 + F4 + F5}.

\subsubsection{Training on Ws and APD and testing on A1, A2, and A3}
% \subsection{Training on Wesad and APD and then testing on our dataset}
Table \ref{tab:cross_gen_train_public_test_own} presents the best results obtained across different feature combinations when classification models were trained on the Ws and APD datasets and tested on A1, A2, and A3. When trained on Ws and tested on A1, A2, and A3, we observed similar evaluation metrics across the activities (see Table \ref{tab:cross_gen_train_public_test_own}, data highlighted in bold). A similar trend was observed when the model was trained on APD and tested on A1, A2, and A3, with AUROCs of 0.64, 0.67, and 0.63, respectively.

Table \ref{app:tab:compact_evaluation_2} in the appendix presents the mean and standard deviation for AUROC and recall percentages for anxious and non-anxious classes across the six train-test combinations (e.g., train Ws, test A1, A2, and A3) for each feature set (data taken from Table \ref{tab:cross_gen_train_public_test_own}). We found the highest AUROC was 0.59, with a recall of 59.43\% for the anxious class and 55.25\% for the non-anxious class, achieved with the feature combination \textit{F1 + F2}.

\section{Discussion}

\begin{table*} [!t]
\scriptsize
\caption{Summary of the best evaluation metric obtained in within-participant classification. Each tuple represents AUROC, Recall (\%) of the anxious class, Recall (\%)  of the non-anxious class. A1: Activity 1, A2: Activity 2, A3: Activity 3.}
\centering

\begin{minipage}{0.40\textwidth}
\subcaption{Within activity (dataset) and within combination of dataset \label{tab:within_summarized}}
\centering
\begin{tabular}{lcc} \toprule
                  & \textbf{ML}          & \textbf{DL}          \\ \midrule
\textbf{A1}       & (0.82, 72.11, 73.68) & (0.77, 68.71, 75.03) \\
\textbf{A2}       & (0.64, 46.90, 80.76) & (0.79, 49.26, 89.48) \\
\textbf{A3}       & (0.82, 56.55, 87.55) & (0.77, 58.29, 84.24)   \\ 
\midrule
\textbf{A1+A2}    & (0.78, 25.44, 94.67) & (0.73, 45.27, 87.15) \\
\textbf{A1+A3}    & (0.77, 58.64, 78.28) & (0.70, 56.30, 74.21) \\
\textbf{A2+A3}    & (0.73, 33.01, 89.79) & (0.69, 34.03, 86.09) \\
\textbf{A1+A2+A3} & (0.73, 25.81, 92.62) & (0.68, 34.81,86.19) \\ \bottomrule
\end{tabular}
\end{minipage}
\hfill
\begin{minipage}{0.58\textwidth}
\subcaption{Cross activity with Fine-tuning \label{tab:cross_finetuning}}
\centering
\begin{tabular}{cccc} \toprule
 \textbf{\textit{Train$\backslash$Fine-tune}} &  \textbf{A1}&  \textbf{A2}& \textbf{A3} \\ \midrule
         \textbf{A1}&  -&  (0.73, 33.33, 84.55)& (0.76, 52.31, 82.16) \\
         \textbf{A2}&  (0.69, 62.50, 70.83)&  -& (0.71, 38.71, 82.26) \\
         \textbf{A3}&  (0.76, 68.75, 66.67)&  (0.72, 33.33, 89.09)& - \\ \bottomrule
\end{tabular}
\end{minipage}

\vspace{0.5cm} % Space between the tables

\begin{minipage}{\textwidth}
\centering
\subcaption{Cross activity (dataset) \label{tab:cross_summarized}}
\begin{tabular}{llcclcclcc} \toprule
\textbf{\textit{$\backslash$Test}}  && \multicolumn{2}{c}{\textbf{A1}} & & \multicolumn{2}{c}{\textbf{A2}} & & \multicolumn{2}{c}{\textbf{A3}}  \\ \cmidrule{1-1} \cmidrule{3-4} \cmidrule{6-7} \cmidrule{9-10}
\textbf{\textit{$\backslash$Train}} && \textit{\textbf{ML}} & \textit{\textbf{DL}}  & & \textit{\textbf{ML}} & \textit{\textbf{DL}}  && \textit{\textbf{ML}} & \textit{\textbf{DL}} \\ \midrule
\textbf{A1}  && \multicolumn{2}{c}{\textbf{-}} & & (0.62,59.70,62.21)   & (0.61,56.11,66.55)    && (0.60,55.48,59.84)   & (0.58, 54.29, 61.32) \\
\textbf{A2}  && (0.61,42.86,78.26)   & (0.60, 35.07, 81.20)  & & \multicolumn{2}{c}{\textbf{-}} &                       & (0.57,38.59,75.58)   & (0.51, 28.41, 78.51) \\
\textbf{A3}  && (0.65,52.70,71.00)   & (0.57, 58.33, 53.71)  & & (0.60,44.11,68.03)   & (0.56, 50.20, 62.25)  && \multicolumn{2}{c}{\textbf{-}}    \\ \bottomrule                
\end{tabular}
\end{minipage}

\label{tab:summarized_result}
\end{table*}

\subsection{Within-participant predictive performance}

\subsubsection{Within-activity analysis:} We conducted a within-activity evaluation to assess the ideal predictive performance of various anxiety detection models. This evaluation approach ensures that the data distribution remains consistent between the training and testing phases, providing a clearer understanding of each model's potential accuracy within a controlled setting. 

We achieved the best AUROC scores of 0.82, 0.79, and 0.82 (see Table~\ref{tab:within_summarized}) for Activities A1, A2, and A3, respectively. The recall \% for the anxious class was 72.11, 49.26, and 56.55 for A1, A2, and A3, respectively. Similarly, the recall \% for the non-anxious class was 73.68, 89.48, and 87.55 for A1, A2, and A3, respectively. The balanced recall for anxious and non-anxious classes in A1 is attributed to the almost equal number of labels in both classes. In contrast, the low recall for anxious and high recall for non-anxious in A2 and A3 may be due to the lower number of anxious labels (approximately 25\% in A2 and 37\% in A3). Balanced data is crucial for training a good classification model; however, obtaining sufficient `mental disorder' labels is challenging in mental health research \cite{pillai2024investigating}. These results suggest that prediction models effectively classify anxious and non-anxious participants when trained on balanced datasets.

% however, obtaining sufficient 'anxious' labels is challenging in mental health research due to the scarcity of anxious participants or anxiety provoking situations. These results suggest that prediction models effectively classify anxious and non-anxious participants when trained on balanced datasets.  

Additionally, we observed an interesting pattern in the ML classifier: among the five classification models tested, C2 (i.e., Random Forest) consistently performed best for almost all feature combinations in A1. Conversely, the C3 (i.e., Decision Tree) performed best for nearly all feature set combinations in A2 and A3. This indicates that both the C3 and its ensemble version, C2, are effective for training prediction models using physiological data for anxiety detection \cite{ancillon2022machine}.

The good performance of the Decision Tree can be due to its ability to model complex relationships in the data through a series of binary decisions. Random Forest enhances this by combining multiple Decision Trees, which helps reduce overfitting and improve generalization to unseen data. The ensemble approach of Random Forest captures a broader range of patterns and interactions within the dataset, contributing to its superior performance in our analysis.

%\hl{Low priority: just check whether other papers have found similar results for DT and RF}

\subsubsection{Cross-activity analysis:} To test the generalizability of the anxiety detection models we treated each activity data as a dataset where we trained and tested on all combinations of our collected datasets comprising of Activities, A1, A2, and A3. This included training on A1 and testing on A2 and A3, training on A2 and testing on A1 and A3, and training on A3 and testing on A1 and A2 for all feature combinations. In total, we trained and tested 930 ML models and 93 DL models. We set a baseline threshold of 50\% since this was a binary classification task.

Our findings indicated that models trained on A1 were generalizable to A2 and A3, surpassing the threshold for all evaluation metrics. The best models showed that they struggled equally to classify both anxious and non-anxious labels, with recall rates for both classes being approximately the same. However, when trained on A2, none of the best models exceeded the recall threshold of 50\% when tested on A1 and A3. This limitation may be due to the low number of anxious samples in the training dataset, which hindered the models' ability to learn the patterns of anxious participants. The highest recall rates achieved were 42.86\% for the anxious class in A1 and 38.59\% in A3. When trained on A3, the model successfully generalized to A1 but faced challenges with A2. The highest recall for the anxious class was found to be 52.70\% in A1 and 50.20\% in A2.

Furthermore, among the best classification models, we observed that the overall Decision Tree outperformed the other algorithms (see Table \ref{tab:cross_gen_train_own_test_public}). There were also notable performances from Logistic Regression and Random Forest, particularly when models were trained on A1 and tested on the other activities. This success may be due to the binary nature of our problem statement, as Logistic Regression is known for effectively learning the representations required for binary classification.

When using the within-participant analysis as a threshold, we found that training on A2 or A3 and testing on A1 achieved AUROCs of only 0.61 and 0.65, respectively—both lower than the 0.82 obtained when both training and testing were conducted on A1. Similarly, training on A1 or A3 and testing on A2 yielded AUROCs of 0.62 and 0.60, which were lower than or equal to the 0.62 obtained from training and testing on A2. However, recall for the anxious class increased by 18\% when the training was done on A1. This can be due to the equally balanced label dataset of A1. When trained on A1 or A2 and tested on A3, the AUROC and recall for the anxious class decreased further. This variation in AUROC and recall suggests that the activity used for training significantly impacts performance, as does the distribution of anxious and non-anxious labels. For instance, A2 had a relatively low number of anxious labels, likely contributing to the decline in anxious class recall when testing A1 and A3.

\subsection{Cross-participant generalization performance}

To evaluate the cross-participant generalizability of our anxiety detection models, we conducted analyses in two settings: (i)  {Cross-activity analysis:} Models were trained on data from either A2 (Group discussion) or A3 (Interview) and tested on data from either Ws (a combination of speech and mathematical tasks) or APD (Speech), and vice versa, and (ii)  {Same-activity analysis:} Models were trained on data from A1 (Speech) and tested on APD (Speech), and vice versa. Our results show that, regardless of the specific training and testing activity, AUROC values consistently ranged from 0.62 to 0.67. Additionally, logistic regression performed well across most feature combinations when the activities matched. In contrast, ensemble learning methods (Decision Tree (C3) and Random Forest (C2)) tended to yield better results when activities differed. Importantly, similar activities did not show a performance advantage over dissimilar activities.

%Next, in line with existing studies exploring models to detect anxiety versus neutral states, we also tested the generalizability of our anxiety detection model for these states. We designated participants with low self-reported anxiety as representing a ``neutral state'' and those with high self-reported anxiety as an ``anxious state''. For validation, we used the WESAD dataset, where 15 participants engaged in both neutral and anxiety-inducing activities. We trained and tested on all combinations of our dataset and the WESAD data. Results showed that models trained on our dataset achieved higher AUROC scores compared to those trained on WESAD. This suggests that anxiety detection models trained on our data generalize effectively to anxiety state detection, underscoring their potential for real-time, just-in-time anxiety detection and further prompting interventions.

\subsection{Machine learning versus DNN for generalizing anxiety detection models}

We used both machine learning (ML) classifiers and a DNN to evaluate the generalizability of anxiety detection models. However, the ML models consistently outperformed the DNN (see Table \ref{tab:cross_summarized}). The only exception occurred in cross-activity training, where the model was trained on A3 and tested on A2, in which case the DNN outperformed the ML models. This suggests that classical ML models may be more effective than deep learning for mental health datasets in both within- and cross-dataset scenarios. 

One advantage of using a DNN is its ability to save model weights and fine-tune on similar datasets if the pre-trained model does not perform well. To evalaute the impact of fine-tuning, we used only the feature combination ``\textit{F1 + F3 + F5}'', as it yielded the best performance for A1, A2, and A3 in within-dataset classification. For example, we first trained the model on A1 using grid search to find the best parameters, then fine-tuned it on A2 and A3. During fine-tuning, we split the dataset (random state set to 42) into training, validation, and test sets in a 70:15:15 ratio. We found that after fine-tuning, the model performed better in some cases while worse in others (see Table \ref{tab:cross_finetuning}).

\subsection{Comparison with previous generalizability studies in mobile sensing}
Generalizability in mobile sensing models, particularly for mental health and behavioral monitoring, is challenging due to diverse demographic, cultural, and behavioral factors \cite{xu2023globem, meegahapola2023generalization, kammoun2023understanding}. In this study, we examined generalizability by testing (1) the same group of participants performing different anxiety-provoking activities, and (2) different participant groups from two countries performing both similar and different anxiety-provoking activities.

For the same group of participants performing activities A1, A2, and A3, the highest AUROC achieved was 0.65, similar to Xu et al. \cite{xu2023globem}, who reported an AUROC of 0.62 when testing generalizability with overlapping participants. In contrast, Meegahapola et al. \cite{meegahapola2023generalization} found that country-agnostic (trained and tested across countries) and country-specific (trained and tested within the same country) models had comparable AUROCs. In our case, however, within-activity models achieved a significantly higher AUROC of 0.82 than 0.61 for cross-activity models. Our findings also support Pillai et al.'s \cite{pillai2024investigating} claim that data similarity enhances transferability is not true. Activity A1 in our dataset was similar to the APD activity in structure and data characteristics, yet cross-activity results between A1 and APD were comparable to those of unrelated activities. Interestingly, while A2 and A3 differed from A1 and APD in activity, they had similar data characteristics (based on t-SNE and OTDD analyses). We also observed that the EDA characteristics of the APD dataset diverged from those of A1, A2, and A3 in our dataset. Despite this, the highest AUROC in generalizability testing included the feature set F5, highlighting how specific features can bolster model performance across datasets, even when broader characteristics differ.

\subsection{Feature importance} 
\begin{figure*}[htbp]
    \centering
    % First subfigure
    \begin{subfigure}[b]{1\linewidth}
        \centering
        \includegraphics[width=\linewidth]{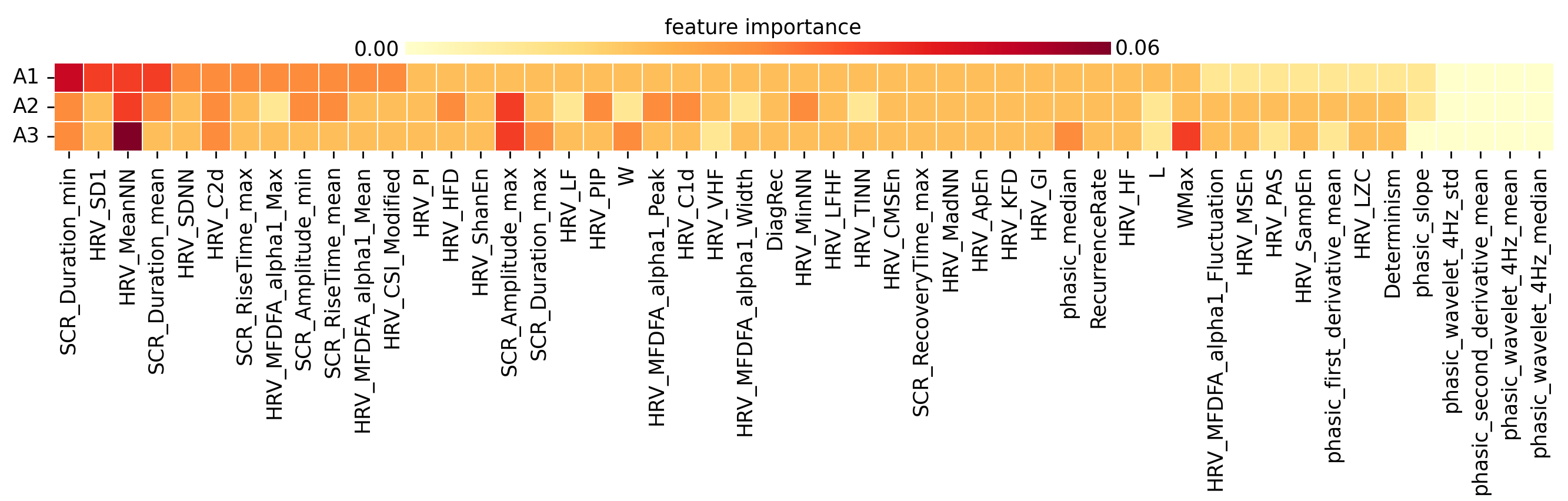}
        \caption{Gini features importance values computed using Random Forest on datasets A1, A2, and A3.}
        \label{fig:gini_own}
    \end{subfigure}
    % Second subfigure
    \begin{subfigure}[b]{1\linewidth}
        \centering
        \includegraphics[width=\linewidth]{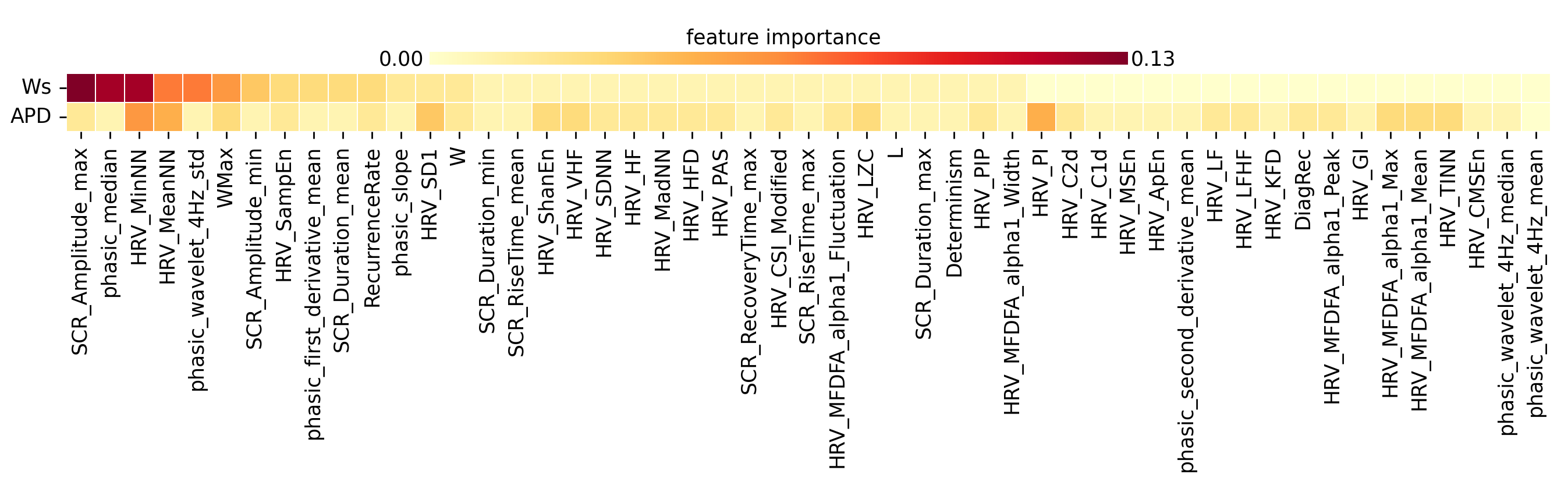}
        \caption{Gini features importance values computed using Random Forest on datasets Ws and APD.}
        \label{fig:gini_public}
    \end{subfigure}
    \caption{Gini features importance values computed using Random Forest on our datasets (i.e., A1, A2, and A3) and publicly available datasets (i.e., Ws and APD).}
    \Description{}
    \label{fig:gini_all}
\end{figure*}

% For Nilesh - Change cheng2022heart to our wearable paper
Figures \ref{fig:gini_own} and \ref{fig:gini_public} present the Gini feature importance values for each activity in our dataset and in publicly available datasets (Ws and APD), respectively. Feature importance values range from 0 to 0.06 in our dataset and from 0 to 0.13 in the publicly available datasets. The results are arranged in descending order, using A1 as the reference in Figure \ref{fig:gini_own} and Ws as the reference in Figure \ref{fig:gini_public}. For our dataset, the top three features in A1 were the minimum SCR duration, SD1, and MeanNN; in A2, the top features were MeanNN, the maximum SCR amplitude, and MinNN; and in A3, they were MeanNN, the maximum SCR amplitude, and WMax. In the publicly available datasets, the top features in Ws included the maximum SCR amplitude, SCR phasic median, and MinNN, while in APD, they were MinNN, MeanNN, and PI. Notably, MeanNN emerged as a significant feature in four out of the five datasets: A1, A2, A3, and APD. This finding aligns with existing research suggesting that MeanNN is one of the key factors in distinguishing between anxiety and non-anxiety states \cite{cheng2022heart}. Additionally, at least one EDA feature was found to be important in four of the five datasets (A1, A2, A3, and Ws). Specifically, the maximum SCR amplitude appeared as a top feature in three datasets (A2, A3, and Ws), underscoring the importance of EDA-extracted features for differentiating between anxiety and non-anxiety. A1 and APD, which both come from student populations but in different countries and share a similar study setup involving a speech activity of comparable duration, identified MeanNN as the only common significant feature.

\subsection{Ethical consideration}
Generalizable anxiety detection models are crucial for real-time or ``just-in-time'' anxiety detection and timely intervention. Developing these models requires capturing physiological data and self-reports from participants in anxiety-inducing situations. While designing experiments to collect physiological data and self-reports in real-time is feasible, it presents challenges, such as obtaining fine-grained self-reports (i.e., annotations) and subjecting participants to repeated anxiety-inducing scenarios. Additionally, ethical and practical constraints limit the frequency with which participants can safely experience such tasks, which in turn impacts the development of generalizable models \cite{mauss2004there}.

To address this, we conducted a study in line with ethical guidelines, where participants completed three different anxiety-inducing tasks. Instead of a single anxiety assessment, participants rated their anxiety after each activity using a fine-grained five-item questionnaire with questions covering the anxiety experienced, emotional arousal, avoidance, excessive worry, and perceived negative evaluation. 
This study design allowed us to collect detailed self-reports and physiological data (e.g., ECG and EDA) across three different anxiety-provoking scenarios, which is essential for developing and evaluating generalizable anxiety detection models. 

\subsection{Limitations}
During data analysis, we identified several limitations in our study that may impact the interpretation of the results.
\begin{enumerate}
    \item The duration of the activities varied, with Activity 2 (A2) lasting 6 minutes, which resulted in a higher number of data windows for A2 compared to Activities 1 and 3 (A1 and A3). A2 involved a group discussion, and the 6-minute duration was chosen to allow each participant approximately 2 minutes to speak on average within a single study session. Ideally, each activity would be of equal duration to enable fair comparisons. However, it is often challenging to standardize activity durations, and accounting for these differences in data analysis can be complex.
    
    \item The group discussion (A2) was the least anxiety-provoking activity, resulting in anxious labels for only 24.23\% of the data. Similarly, A3 showed just 37.84\% of anxious labels. The low proportion of anxious labels likely impacted the training of the machine learning models, limiting their ability to learn patterns specific to anxious participants. However, in real-world settings, individuals react differently to various situations, with some activities provoking higher levels of anxiety while others may provoke lower or even no anxiety.
    
    % \item \hl{check the fit} A substantial amount of research has focused on detecting baseline and emotional arousal states. In mental health studies, researchers have also developed predictive models to classify participants' baseline and TSST states. Typically, these studies involve participants completing baseline and TSST activities, after which each phase is labeled as 'anxious' or 'non-anxious.' While this approach is effective for detecting emotional arousal, it may not be ideal for anxiety detection, as it assumes that the baseline is always non-anxious and the TSST is always anxious.
    % \item \hl{check the fit}Moreover, physiological responses to anxiety may vary across activities; one activity may provoke anxiety for a participant, while another might not. Ideally, a personalized model would require participants to undergo multiple anxiety-inducing activities, which is impractical. However, if the same group of participants undergoes even a smaller number of anxiety-provoking tasks, it can yield a model with characteristics close to a personalized one due to the similarity in data patterns. OR .So, instead of focusing on a personalized anxiety detection model, a generalized anxiety detection model can be explored.
\end{enumerate}

\subsection{Implications}
\begin{enumerate}
    \item \textit{Just-in-time anxiety detection}: Advances in wearable technology, alongside decreasing costs, have made wearables highly accessible and seamlessly integrated into daily life for many. Our within-activity analysis indicates that anxious individuals can be identified during anxiety-provoking activities, allowing for real-time determination of their current emotional state. Additionally, cross-activity results suggest that the ML models can detect anxiety even when an individual is engaged in a different type of activity. This highlights the model’s potential for real-time (just-in-time) anxiety detection, supporting timely interventions. Our work could further be extended to interpret user context, identifying specific activities or scenarios that trigger anxiety and enabling targeted and scenario-specific interventions.
    \item \textit{Contribution to mental health research}: Our research contributes to mental health studies by evaluating the generalizibility of anxiety detection models. These models can help identify individuals with anxiety using physiological data. By providing an accessible screening tool, it may help bridge treatment gaps and reduce pressure on limited mental health resources, particularly in low- and lower-middle-income countries where mental health care is often underfunded.
    
    % Our dataset is unique compared to publicly available datasets because it includes a variety of physiological markers collected across three distinct anxiety-provoking activities. This dataset provides valuable opportunities for the UbiComp community to explore questions related to (i) generalizability, (ii) physiological analysis, and (iii) behavioral analysis. It can serve as a baseline for evaluating other methods and datasets, supporting replication and validation of results in future research.
\end{enumerate}
\subsection{Future work}
In this work, we assessed the generalizability of an anxiety detection models using machine learning, given the small dataset size and the interpretability advantages of machine learning models over deep learning. However, deep learning approaches may also be worth exploring, particularly as we did not achieve satisfactory AUROC values in our generalizability analysis. Additionally, transfer learning could be explored by training the model on one dataset, fine-tuning it on another, and then testing its performance. Finally, feature fusion could be investigated, as our data include two different modalities, i.e., ECG and EDA, which may provide complementary information for anxiety detection that machine learning models were unable to capture. Moreover, anxiety-inducing activity duration and window size selection can be studied to see their effect on the anxiety detection model.

\section{Conclusion}
In this work, we assessed the generalizability of an anxiety detection models trained on biobehavioral data collected through wearable sensors from participants engaged in anxiety-inducing activities. Given the importance of generalizability in just-in-time (JIT) anxiety detection for timely intervention, we aimed to determine whether a model trained on one anxiety-inducing activity could effectively predict anxiety when participants performed different anxiety-inducing activities. To explore this, we designed a study where the same pool of participants engaged in three distinct anxiety-provoking tasks. Using our study data, we examined generalizability within participants for both within-activity and cross-activity scenarios. We extended this exploration to publicly available datasets, which included activities both similar and dissimilar to those in our study, allowing us to analyze generalizability across participants and activities. Our results showed AUROC values ranging from 0.62 to 0.73. Notably, we observed no considerable effect of activity type on model performance, as AUROC values remained similar regardless of whether the model was applied within or across participants performing the same or different activities.

\bibliographystyle{ACM-Reference-Format}
\bibliography{sample-base}

\appendix

\onecolumn
\section{Appendix}

\setcounter{table}{0}
\setcounter{figure}{0}
\renewcommand{\thetable}{A.\arabic{table}}
\renewcommand{\thefigure}{A.\arabic{figure}}

\begin{table}[h]
\centering
% \small
\caption{Within-activity (within-dataset) classification evaluation metrics with different combinations of feature sets for Activity 1, Activity 2, and Activity 3 using deep neural network (DNN). Each tuple represents AUROC, Recall (\%) of the anxious class, Recall (\%)  of the non-anxious class. Column-wise bold tuples represent the best AUROC performance observed during the within-dataset evaluation. In cases of AUROC ties, the feature set with the highest Anxious Recall was selected.
}
\label{app:tab:DNN_within_dataset}
\begin{tabular}{@{}lccc@{}}
\toprule
\textbf{Feature Set}            & \textbf{Activity 1}       & \textbf{Activity 2}       & \textbf{Activity 3}       \\ \midrule
\textit{Baseline}               & (0.5,0.5,0.5,-)           &(0.5,0.5,0.5,-)            & (0.5,0.5,0.5,-) \\ \midrule
\textit{F1}                     & (0.75, 0.72, 0.66)
& (0.72, 0.32, 0.87)
& (0.71, 0.5, 0.8)
\\
\textit{F2}                     & (0.53, 0.49, 0.58)
& (0.58, 0.11, 0.94)
& (0.52, 0.27, 0.75)
\\
\textit{F3}                     & (0.71, 0.67, 0.64)
& (0.71, 0.42, 0.88)
& (0.67, 0.47, 0.77)
\\
\textit{F4}                     & (0.58, 0.56, 0.52)
& (0.57, 0.27, 0.83)
& (0.55, 0.41, 0.67)
\\
\textit{F5}                     & (0.64, 0.61, 0.64)
& (0.58, 0.25, 0.81)
& (0.61, 0.4, 0.74)
\\ \midrule
\textit{F1 + F2}                & (0.64, 0.57, 0.64)
& (0.74, 0.4, 0.86)
& (0.64, 0.44, 0.7)
\\
\textit{F1 + F3}                & (0.75, 0.72, 0.68)
& (0.75, 0.45, 0.89)
& (0.72, 0.49, 0.81)
\\
\textit{F1 + F4}                & (0.67, 0.66, 0.57)
& (0.62, 0.29, 0.83)
& (0.61, 0.47, 0.72)
\\
\textit{F1 + F5}                & (0.74, 0.68, 0.66)
& (0.68, 0.38, 0.83)
& (0.75, 0.6, 0.78)
\\
\textit{F2 + F3}                & (0.73, 0.65, 0.67)
& (0.71, 0.39, 0.86)
& (0.64, 0.43, 0.76)
\\
\textit{F2 + F4}                & (0.63, 0.64, 0.56)
& (0.59, 0.3, 0.84)
& (0.58, 0.42, 0.64)
\\
\textit{F2 + F5}                & (0.66, 0.63, 0.63)
& (0.62, 0.32, 0.83)
& (0.67, 0.49, 0.75)
\\
\textit{F3 + F4}                & (0.67, 0.68, 0.6)
& (0.69, 0.37, 0.86)
& (0.68, 0.54, 0.73)
\\
\textit{F3 + F5}                & (0.75, 0.7, 0.72)
& (0.75, 0.46, 0.86)
& (0.74, 0.54, 0.79)
\\
\textit{F4 + F5}                & (0.7, 0.73, 0.59)
& (0.64, 0.38, 0.83)
& (0.68, 0.51, 0.71)
\\ \midrule
\textit{F1 + F2 + F3}           & (0.77, 0.68, 0.71)
& (0.75, 0.44, 0.89)
& (0.68, 0.48, 0.81)
\\
\textit{F1 + F2 + F4}           & (0.69, 0.68, 0.59)
& (0.66, 0.38, 0.84)
& (0.64, 0.49, 0.7)
\\
\textit{F1 + F2 + F5}           & (0.73, 0.64, 0.72)
& (0.74, 0.43, 0.86)
& (0.72, 0.56, 0.75)
\\
\textit{F1 + F3 + F4}           & (0.72, 0.69, 0.62)
& (0.7, 0.38, 0.9)
& (0.7, 0.5, 0.74)
\\
\textit{F1 + F3 + F5}           & (0.76, 0.72, 0.71)
& (0.78, 0.48, 0.89)
& (0.76, 0.56, 0.82)
\\
\textit{F1 + F4 + F5}           & (0.75, 0.72, 0.66)
& (0.67, 0.44, 0.84)
& (0.73, 0.52, 0.76)
\\
\textit{F2 + F3 + F4}           & (0.69, 0.7, 0.63)
& (0.69, 0.41, 0.87)
& (0.68, 0.53, 0.75)
\\
\textit{F2 + F3 + F5}           & (0.75, 0.64, 0.72)
& (0.77, 0.47, 0.89)
& (0.73, 0.53, 0.79)
\\
\textit{F2 + F4 + F5}           & (0.72, 0.67, 0.64)
& (0.68, 0.43, 0.86)
& (0.68, 0.5, 0.73)
\\ \midrule
\textit{F3 + F4 + F5}           & (0.72, 0.69, 0.68)
& (0.74, 0.45, 0.88)
& (0.7, 0.54, 0.75)
\\
\textit{F1 + F2 + F3 + F4}      & (0.71, 0.68, 0.6)
& (0.71, 0.39, 0.88)
& (0.72, 0.54, 0.76)
\\
\textit{F1 + F2 + F3 + F5}      & (0.72, 0.64, 0.67)
& (0.79, 0.53, 0.89)
& (0.76, 0.54, 0.82)
\\
\textit{F1 + F2 + F4 + F5}      & (0.77, 0.72, 0.68)
& (0.69, 0.43, 0.84)
& (0.72, 0.54, 0.73)
\\
\textit{F1 + F3 + F4 + F5}      & (0.75, 0.73, 0.68)
& (0.76, 0.49, 0.89)
& (0.74, 0.55, 0.78)
\\
\textit{F2 + F3 + F4 + F5}      & (0.73, 0.71, 0.68)
& (0.77, 0.45, 0.9)
& (0.72, 0.58, 0.76)
\\ \midrule
\textit{F1 + F2 + F3 + F4 + F5} & (0.77, 0.73, 0.72)
& (0.78, 0.52, 0.89)
& (0.75, 0.56, 0.75)
\\ \bottomrule
\end{tabular}
\end{table}

\begin{table}[htbp]
\centering
% \scriptsize
\caption{AUROC, Recall (\%) of the anxious class, and Recall (\%) of the non-anxious class averaged across six train-test combinations. Each row represent row-wise averages of Table \ref{tab:in_detail_generalizability_evaluation_metric} metrics.}
\label{app:tab:compact_evaluation}
\begin{tabular}{@{}llclcc@{}}
\toprule
\multirow{2}{*}{\textbf{Features}}      && \multirow{2}{*}{\textbf{AUROC}}  && \multicolumn{2}{c}{\textbf{Recall (\%)}} \\ \cmidrule{5-6}
 &&    &&   \textit{\textbf{Anxious}} & \textit{\textbf{Non-anxious}}  \\ \midrule
\textit{Baseline} && 0.50 (0.00) && 0.50 (0.00)& 0.50 (0.00) \\ \midrule
F1                      && 0.55 (0.02) && 42.95 (4.78) & 65.61 (2.21)
 \\
F2                      && 0.50 (0.01) && 38.06 (5.35) & 62.86 (5.63)
 \\
F3                      && 0.54 (0.02) && 41.00 (6.11) & 67.80 (6.12)
 \\
F4                      && 0.51 (0.02) && 39.79 (8.73) & 61.56 (4.69)
 \\
F5                      && 0.51 (0.02) && 33.47 (8.76) & 67.94 (8.25)
 \\ \midrule
 mean ($\sigma$)& & 0.52 (0.03) & & 39.05 (6.75)& 65.15 (5.53)
\\ \midrule
F1 + F2                 && && 40.48 (3.45) & 68.81 (1.49)
 \\
F1 + F3                 && 0.57 (0.00) && 43.38 (5.49) & 67.48 (4.80)
 \\
F1 + F4                 && 0.54 (0.01) && 44.49 (4.96) & 64.33 (4.70)
 \\
F1 + F5                 && 0.56 (0.01) && 44.74 (4.40) & 65.82 (3.67)
 \\
F2 + F3                 && 0.53 (0.01) && 40.05 (4.12) & 65.36 (2.74)
 \\
F2 + F4                 && 0.53 (0.01) && 43.51 (7.33) & 61.90 (6.51)
 \\
F2 + F5                 && 0.51 (0.02) && 37.69 (9.58) & 64.54 (7.75)
 \\
F3 + F4                 && 0.55 (0.01) && 43.09 (7.34) & 64.83 (6.57)
 \\
F3 + F5                 && 0.54 (0.03) && 43.43 (7.60) & 62.82 (2.53)
 \\
F4 + F5                 && 0.52 (0.01) && 40.13 (9.03) & 63.10 (7.99)
 \\ \midrule
 mean ($\sigma$)& & 0.54 (0.02) & & 42.10 (5.96)& 64.90 (4.87)
\\ \midrule
F1 + F2 + F3            && && 42.53 (5.55) & 68.54 (5.10)
 \\
F1 + F2 + F4            && 0.55 (0.01) && 44.63 (7.91) & 64.32 (6.29)
 \\
F1 + F2 + F5            && 0.55 (0.01) && 43.74 (3.87) & 65.86 (3.14)
 \\
F1 + F3 + F4            && \textbf{0.59 (0.01)} && 47.75 (5.54) & 67.61 (4.59)
 \\
F1 + F3 + F5            && 0.53 (0.01) && 42.28 (4.41) & 64.17 (3.18)
 \\
F1 + F4 + F5            && 0.54 (0.02) && 46.70 (3.31) & 62.08 (3.26)
 \\
F2 + F3 + F4            && 0.56 (0.01) && 43.35 (7.43) & 66.50 (4.85)
 \\
F2 + F3 + F5            && 0.53 (0.01) && 40.66 (4.87) & 63.65 (3.04)
 \\
F2 + F4 + F5            && 0.52 (0.02) && 39.96 (3.54) & 63.61 (2.31)
 \\
F3 + F4 + F5            && 0.53 (0.01) && 41.96 (4.43) & 62.25 (2.97)
 \\ \midrule
 mean ($\sigma$)& & 0.55 (0.03) & & 43.35 (5.00)& 64.86 (3.97)
\\ \midrule
F1 + F2 + F3 +   F4     && && 46.93 (6.28) & 67.32 (2.59)
 \\
F1 + F2 + F3 + F5       && 0.55 (0.01) && 40.46 (1.82) & 68.42 (2.10)
 \\
F1 + F2 + F4 + F5       && 0.54 (0.01) && 43.79 (6.28) & 63.81 (4.36)
 \\
F1 + F3 + F4 + F5       && 0.57 (0.02) && 46.24 (7.96) & 65.72 (6.09)
 \\
F2 + F3 + F4 + F5       && 0.52 (0.01) && 42.23 (4.54) & 61.34 (2.57)
 \\ \midrule
 mean ($\sigma$)& & 0.55 (0.03) & & 43.93 (5.48)& 65.32 (4.17)
\\ \midrule
F1 + F2 + F3 + F4 + F5  && && 46.26 (7.55) & 65.67 (6.76)
 \\ \bottomrule
\end{tabular}
\end{table}

\begin{table}[htbp]
    \centering
    \begin{minipage}{0.47\textwidth}
        \centering
        \scriptsize
        \caption{AUROC, Recall (\%) of the anxious class, and Recall (\%) of the non-anxious class averaged across six train-test combinations (Training on A1, A2, A3 and Test on Ws, APD). Each row represent row-wise averages of Table  \ref{tab:cross_gen_train_own_test_public}.}  
        \label{app:tab:compact_evaluation_1}
        \begin{tabular}{@{}llclcc@{}}
            \toprule
            \multirow{2}{*}{\textbf{Features}}      && \multirow{2}{*}{\textbf{AUROC}}  && \multicolumn{2}{c}{\textbf{Recall (\%)}}  \\ \cmidrule{5-6}
            &&    &&   \textit{\textbf{Anxious}} & \textit{\textbf{Non-anxious}} \\ \midrule
            \textit{Baseline} && 0.5 (0.0) && 0.5 (0.0) & 0.5 (0.0) \\ \midrule
            F1 && 0.56 (0.04) && 34.81 (5.08) & 73.35 (2.71) \\
            F2 && 0.52 (0.01) && 39.02 (0.93) & 65.54 (0.65) \\
            F3 && 0.51 (0.00) && 35.21 (4.53) & 67.10 (3.88) \\
            F4 && 0.55 (0.04) && 43.73 (4.37) & 63.70 (0.18) \\
            F5 && 0.52 (0.05) && 39.57 (18.62) & 67.17 (9.76) \\ \midrule
            F1 + F2 && 0.55 (0.02) && 38.94 (3.44) & 67.53 (0.07) \\
            F1 + F3 && 0.54 (0.04) && 40.84 (3.50) & 66.22 (3.21) \\
            F1 + F4 && 0.54 (0.02) && 38.71 (6.77) & 70.30 (6.92) \\
            F1 + F5 && 0.57 (0.02) && 34.85 (19.25) & 76.63 (20.83) \\
            F2 + F3 && 0.52 (0.03) && 37.06 (1.90) & 65.01 (1.18) \\
            F2 + F4 && 0.56 (0.05) && 36.70 (3.51) & 71.27 (8.70) \\
            F2 + F5 && 0.54 (0.07) && 37.39 (18.42) & 70.28 (6.71) \\
            F3 + F4 && 0.53 (0.01) && 41.58 (0.33) & 64.57 (0.04) \\
            F3 + F5 && 0.54 (0.02) && 41.34 (13.24) & 67.37 (9.33) \\
            F4 + F5 && 0.57 (0.01) && 40.64 (10.82) & 70.47 (8.67) \\ \midrule
            F1 + F2 + F3 && 0.54 (0.03) && 41.49 (6.30) & 67.02 (1.11) \\
            F1 + F2 + F4 && 0.53 (0.06) && 36.71 (3.11) & 66.65 (4.00) \\
            F1 + F2 + F5 && 0.52 (0.04) && 31.17 (24.46) & 73.89 (17.92) \\
            F1 + F3 + F4 && 0.51 (0.03) && 41.83 (0.81) & 59.03 (6.01) \\
            F1 + F3 + F5 && 0.59 (0.02) && 41.74 (10.69) & 73.23 (6.07) \\
            F1 + F4 + F5 && 0.60 (0.04) && 48.69 (9.13) & 70.20 (13.77) \\
            F2 + F3 + F4 && 0.53 (0.00) && 39.58 (3.58) & 65.92 (0.32) \\
            F2 + F3 + F5 && 0.55 (0.08) && 37.90 (20.09) & 72.36 (4.85) \\
            F2 + F4 + F5 && 0.57 (0.01) && 53.64 (4.32) & 56.33 (2.56) \\
            F3 + F4 + F5 && 0.58 (0.01) && 42.19 (11.70) & 72.27 (10.85) \\ \midrule
            F1 + F2 + F3 + F4 && 0.53 (0.01) && 33.82 (0.55) & 69.88 (4.96) \\
            F1 + F2 + F3 + F5 && 0.59 (0.01) && 37.86 (15.38) & 77.88 (15.53) \\
            F1 + F2 + F4 + F5 && 0.62 (0.03) && 47.65 (7.95) & 73.10 (16.40) \\
            F1 + F3 + F4 + F5 && 0.59 (0.03) && 41.86 (12.61) & 73.64 (14.25) \\
            F2 + F3 + F4 + F5 && 0.60 (0.02) && 45.94 (14.78) & 70.40 (10.76) \\ \midrule
            F1 + F2 + F3 + F4 + F5 && 0.57 (0.01) && 44.48 (8.02) & 67.09 (3.52) \\ \bottomrule
        \end{tabular}
    \end{minipage}%
    \hfill
    \begin{minipage}{0.47\textwidth}
        \centering
        \scriptsize
        \caption{AUROC, Recall (\%) of the anxious class, and Recall (\%) of the non-anxious class averaged across six train-test combinations (Training on Ws, APD and Test on A1, A2, A3). Each row represent row-wise averages of  Table \ref{tab:cross_gen_train_public_test_own}.}
        \label{app:tab:compact_evaluation_2}
        \begin{tabular}{@{}llclcc@{}}
            \toprule
            \multirow{2}{*}{\textbf{Features}}      && \multirow{2}{*}{\textbf{AUROC}}  && \multicolumn{2}{c}{\textbf{Recall (\%)}}  \\ \cmidrule{5-6}
            &&    &&   \textit{\textbf{Anxious}} & \textit{\textbf{Non-anxious}} \\ \midrule
            \textit{Baseline} && 0.5 (0.0) && 0.5 (0.0) & 0.5 (0.0) \\ \midrule
            F1 && 0.57 (0.02) && 55.61 (6.66) & 55.14 (4.34) \\
            F2 && 0.52 (0.02) && 50.44 (3.09) & 53.76 (0.81) \\
            F3 && 0.53 (0.01) && 47.12 (2.10) & 59.11 (2.97) \\
            F4 && 0.58 (0.04) && 59.18 (2.88) & 56.44 (8.57) \\
            F5 && 0.53 (0.09) && 26.30 (11.18) & 77.03 (5.48) \\ \midrule
            F1 + F2 && 0.59 (0.01) && 59.43 (3.85) & 55.22 (0.01) \\
            F1 + F3 && 0.56 (0.01) && 55.59 (1.60) & 55.01 (2.68) \\
            F1 + F4 && 0.54 (0.03) && 50.34 (2.89) & 56.07 (2.79) \\
            F1 + F5 && 0.58 (0.02) && 48.51 (6.37) & 63.44 (6.85) \\
            F2 + F3 && 0.53 (0.03) && 44.41 (4.92) & 61.16 (6.45) \\
            F2 + F4 && 0.57 (0.02) && 58.50 (9.22) & 53.03 (2.09) \\
            F2 + F5 && 0.54 (0.06) && 49.42 (13.41) & 56.90 (4.27) \\
            F3 + F4 && 0.54 (0.01) && 50.14 (2.34) & 55.68 (3.29) \\
            F3 + F5 && 0.56 (0.03) && 52.88 (6.98) & 56.03 (0.32) \\
            F4 + F5 && 0.58 (0.05) && 49.72 (9.73) & 61.78 (3.32) \\ \midrule
            F1 + F2 + F3 && 0.56 (0.03) && 46.55 (7.82) & 61.22 (4.96) \\
            F1 + F2 + F4 && 0.58 (0.06) && 62.03 (7.46) & 50.74 (0.95) \\
            F1 + F2 + F5 && 0.59 (0.01) && 51.98 (11.96) & 61.56 (11.09) \\
            F1 + F3 + F4 && 0.57 (0.04) && 46.79 (14.19) & 61.64 (6.47) \\
            F1 + F3 + F5 && 0.58 (0.02) && 52.85 (3.82) & 59.96 (4.19) \\
            F1 + F4 + F5 && 0.56 (0.04) && 39.58 (31.63) & 67.91 (24.79) \\
            F2 + F3 + F4 && 0.56 (0.04) && 48.03 (9.82) & 61.38 (5.42) \\
            F2 + F3 + F5 && 0.55 (0.02) && 51.31 (5.71) & 56.76 (1.10) \\
            F2 + F4 + F5 && 0.59 (0.03) && 46.84 (5.07) & 64.77 (1.09) \\
            F3 + F4 + F5 && 0.55 (0.03) && 43.59 (6.20) & 61.22 (1.73) \\ \midrule
            F1 + F2 + F3 + F4 && 0.57 (0.04) && 47.34 (19.40) & 62.47 (9.97) \\
            F1 + F2 + F3 + F5 && 0.57 (0.03) && 49.38 (9.03) & 59.90 (3.53) \\
            F1 + F2 + F4 + F5 && 0.58 (0.06) && 43.73 (21.48) & 67.17 (12.00) \\
            F1 + F3 + F4 + F5 && 0.55 (0.05) && 31.32 (8.64) & 73.81 (2.86) \\
            F2 + F3 + F4 + F5 && 0.57 (0.04) && 51.92 (7.26) & 60.46 (4.87) \\ \midrule
            F1 + F2 + F3 + F4 + F5 && 0.53 (0.04) && 27.84 (7.51) & 74.26 (3.82) \\ \bottomrule
        \end{tabular}
    \end{minipage}
\end{table}
\end{document}